\documentclass[useAMS,usenatbib]{mn2e}

\newcommand{\ltsimeq}{\raisebox{-0.6ex}{$\,\stackrel 
        {\raisebox{-.2ex}{$\textstyle <$}}{\sim}\,$}}

\begin{document}

\title[Two 100 Mpc-scale structures in the 3-D distribution of radio galaxies and their implications]
{Two 100 Mpc-scale structures in the 3-D distribution of radio galaxies and their implications}

\author[Brand et al.]{Kate Brand$^{1\star}$,Steve Rawlings$^{1}$,Gary J. Hill$^{3}$,Mark Lacy$^{2}$,Ewan Mitchell$^{1}$,Joe Tufts$^{3}$ \\
$^1$Astrophysics, Department of Physics, Keble Road, Oxford OX1 3RH, UK \\
$^2$SIRTF Science Center; MS-220-6; California Institute of Technology, 1200 E. California Boulevard, Pasadena, CA 91125\\
$^3$McDonald Observatory and Department of Astronomy, University of Texas at Austin, RLM 15.308, Austin, TX 78712, USA
}
\maketitle

\begin{abstract}
\noindent

We present unequivocal evidence for a huge ($\approx$ 80 $\times$ 100 $\times$ 100 $\rm {Mpc}^3$) super-structure at redshift $z=0.27$ in the 3-D distribution of radio galaxies from the TONS08 sample, confirming tentative evidence for such a structure from the 7C redshift survey (7CRS). A second, newly discovered super-structure is also less securely found at redshift 0.35 (of dimensions $\approx$ 100 $\times$ 100 $\times$ 100 $\rm{Mpc}^3$). We present full observational details on the TONS08 sample which was constructed to probe structures in the redshift range $0\ltsimeq z\ltsimeq0.5$ by matching NVSS sources with objects in APM catalogues to obtain a sample of optically bright (E$\approx$R $\le$ 19.83), radio faint (1.4-GHz flux density S$_{1.4} \ge$ 3 mJy) radio galaxies in the same 25 deg$^{2}$ area as part-II of the 7CRS. Out of the total sample size of 84 radio galaxies, at least 25 are associated with the two $\approx$100 Mpc-scale super-structures. We use quasi-linear structure formation theory to estimate the number of such structures expected in the TONS08 volume if the canonical value for radio galaxy bias is assumed. Under this assumption, the structures represent $\approx$ 4-5$\sigma$ peaks in the primordial density field and their expected number is low ($\sim 10^{-2}-10^{-4}$). Because the TONS08 survey was designed to follow up a previously, if tentatively, identified super-structure in the 7CRS, the probability of finding two super-structures in TONS08 is uncertain but, assuming that the tentative detection of the $z$=0.35 super-structure is real, must lie between $\sim 10^{-4}$ and $\sim 10^{-5}$ depending on how representative the TONS08 region proves to be. We show that similar structures (with similarly low probabilities) are also found in previous radio galaxy redshift surveys but their significance has not been fully appreciated because they have been traced by very small numbers of radio galaxies. Fortunately, there are several plausible explanations (many of which are testable) for these low probabilities in the form of potential mechanisms for boosting the bias on large scales. These include: the association of radio galaxies with highly biased rich clusters in super-structures, enhanced triggering by group/group mergers, and enhanced triggering and/or redshift space distortion in collapsing systems as the growth of super-structures moves into the non-linear regime. 

\end{abstract}


\footnotetext{$^{\star}$Email: brand@astro.ox.ac.uk}

\section{INTRODUCTION}
\label{sec:intro}

Super-structures are the largest known structures in the Universe. It seems they have evolved from the rare ($>$3$\sigma$) peaks in the initial density field at recombination \citep{hna}. They are identified in the local Universe as aggregates of clusters of galaxies, superclusters or large ( $\sim$ 100 Mpc x 100 Mpc) but thin ($\sim$ 10 Mpc) sheets of galaxies such as the Great Wall \citep{gh}. Previous large-scale surveys suggest that the power spectrum of galaxy density has considerable strength on scales up to at least 100 Mpc \citep{per}. Cosmological biasing predicts that large-scale dark-matter fluctuations help push small-scale fluctuations over the density contrast required for gravitational collapse \citep{kai}, meaning that aggregates of rich clusters are expected to trace super-structures in the same way that aggregates of galaxies trace rich clusters. 
Study of these structures is difficult due to the huge areas of sky needed to survey them. Radio galaxies allow us to cover larger areas efficiently because they are biased tracers of the mass \citep{pd}. The bias factor quantifies how well different types of galaxies trace the underlying mass. Relative bias factors are also used to determine how well one population of galaxies traces another. \citet{pd} find a relative bias factor of 1.5 for low redshift radio galaxies with respect to that of optical galaxies (compared to 3.5 for clusters). Recent results from joint 2dFGRS and CMB data \citep{lah} and the analysis of the bi-spectrum of the 2dFGRS \citep{ver} show that the bias parameter for optically selected galaxies is $b\approx1$ with respect to the underlying dark matter, which is consistent with no biasing on scales of tens of Mpc. There have been suggestions of luminosity-dependent biasing in which intrinsically brighter galaxies are more biased. This would not be surprising as larger, brighter galaxies may preferentially form in high density regions therefore being ``born'' biased. Radio galaxies are almost exclusively found to be associated with massive elliptical galaxies, which themselves tend to reside in clusters. It is therefore not surprising that radio galaxies are biased tracers of mass in the local Universe.

It was only with the advent of radio surveys such as the NRAO VLA Sky Survey (NVSS; \citealt{con}) and the Faint Images of the Radio Sky at Twenty-centimeters survey (FIRST; \citealt{bwh}) which now cover most of the sky down to sensitivities of a few mJy at 1.4 GHz, that we can compile the large area samples of faint radio galaxies that we need to study large-scale structure. Radio surveys also have the advantage of suffering no dust obscuration, unlike their optical counterparts. They therefore provide very clean source selection and probe to large redshifts (for NVSS, the median redshift is $\approx$ 1; \citealt{con}). Signatures of clustering and large-scale structure are now well established in the projected distribution of these sources (e.g. \citet{bw} and references therein). 

New galaxy redshift surveys such as the 2dF Galaxy Redshift Survey (2dFGRS) \citep{col} and the Sloan Digital Sky Survey (SDSS)  \citep{sto} which exploit multi-fibre spectrographs to obtain thousands of redshifts can be used to match radio positions with galaxies and hence obtain the largest redshift samples of radio galaxies with redshifts yet. This allows studies in 3-D which have begun with early subsamples of survey data. \citet{sad} and \citet{mag} have matched the 2dFGRS with NVSS and FIRST positions respectively to measure the Radio Luminosity Function (RLF) at low redshift. These studies reveal two distinct populations of radio galaxies at faint ($S_{\rm{1.4}}\sim$ 1 mJy) levels: classical radio galaxies and quasars powered by Active Galactic Nuclei (AGN) and star-forming galaxies in which the radio emission arises through synchrotron emission from relativistic electrons accelerated by supernovae. An increasing fraction of the radio population is identified with starbursts at decreasing flux densities. Due to their lower radio luminosities, these objects are typically at lower redshift than the radio AGN. Because of its flux-density limit, almost all NVSS radio galaxies at $z>0.1$ are associated with AGN rather than starbursts. 

This paper is concerned with a new study of the 3-D distribution of radio galaxies at NVSS flux density levels. It will be structured in the following way. In Sec. 2 we describe the selection techniques used in the Texas-Oxford NVSS Structure (TONS) redshift survey of radio galaxies: this study is based on the results of this survey in the TONS08 region chosen to overlap with part of the 7CRS in which tentative evidence for a super-structure has already been found \citep{rhw}. We present the observational data on TONS08 in Sec. 3. Sec. 4 describes analysis of the resulting redshift distribution including discussion on the significance of redshift peaks, methods of assigning an overdensity to the distribution and methods for calculating how rare these overdensities must be given an assumed bias. We also quantify the geometry of the super-structure using a principal components analysis (PCA). Sec. 5 is a discussion and the main points of the paper are summarised in the conclusions in Sec. 6.  

Unless otherwise stated, we assume a spatially flat $\Lambda$CDM Universe throughout the paper with the following values for the cosmological parameters:
Hubble constant: $H_{0}=70~ {\rm km~s^{-1}Mpc^{-1}}$; $h=H_{0}/100=0.7$; matter density parameter at $z$=0: $\Omega_ {\rm M}(0)=0.3$; vacuum density parameter at z=0: $\Omega_ {\Lambda}(0)=0.7$; rms density variation in 8 Mpc spheres: $\sigma_{8h^{-1}}=0.94$; shape parameter $\Gamma=0.21$; and index of the power spectrum n=1. 

\section{SAMPLE SELECTION}
\label{sec:selection}

The TONS08 (Texas-Oxford NVSS Structure 08$^h$ region) survey is in one of the areas covered by the 7CRS \citep{wil02} and the TexOx-1000 (TOOT) survey \citep{hr}. It covers the region: $08^h10^m20^s\le$ RA $\le08^h29^m20^s$ and $24^\circ10^\prime00^{\prime\prime}\le$ DEC $\le29^\circ30^\prime00^{\prime\prime}$ (J2000). Unlike the low-frequency selected 7CRS and TOOT, the TONS08 survey is selected at 1.4 GHz from the NVSS. For objects of typical spectral index, it therefore goes to fainter radio flux densities than TOOT (which has a 151 MHz flux density $S_{\rm{151}}$ limit of 100 mJy, corresponding to $S_{\rm{1.4}}\approx$20 mJy for radio spectral index\footnote{$\alpha$ is the spectral index for radio sources where the radio flux density $S_{\rm{\nu}} \propto \nu^{-\alpha}$, where $\nu$ is the observing frequency}$\alpha\approx$0.8). In addition, TONS08 has an optical magnitude limit ($E\approx R\approx$19.5) imposed on it.  
TONS08 was specifically designed to seek further evidence for a giant super-structure found in this region of sky by analysis of the redshift distribution of in the 7C-II survey \citep{rhw}. The redshift distribution in the 7C-II survey revealed a prominent spike at $z$ $\approx$ 0.25 of 5 radio galaxies from the 7CRS. Two further radio-loud objects at this redshift were also found: one was just below the 151 MHz flux density ($S_{\rm{151}}$) limit (7C0811+2838), the other was just outside the RA limit (NV0840+2805). 

By going to fainter radio flux densities ($S_{\rm{1.4}}\ge$3 mJy) and optical limits ($E\approx R\approx$19.5), the TONS08 sample was optimised for looking at clustering of objects in a region of moderate (0 $\ltsimeq$ $z$ $\ltsimeq$ 0.5) redshifts. Table.~\ref{tab:samples} gives a summary of the various radio galaxy samples that are used or referred to in this paper.

\scriptsize
\begin{table*}
\begin{center}
\begin{tabular}{lllllll}
\hline\hline
Survey & Type & radio flux density limits & optical Mag limits & colour cut & area (sr.) & ref. \\
\hline
TONS08 & $ROz$ & $S_{1.4}>$3 mJy & $E<$19.83 & $B-R >$1.8 & 0.00688 & This paper\\
TONS12  & $ROz$ & $S_{1.4}>$3 mJy  & $E\approx R<$19.5  & $B-R >$1.8 & 0.00496 & Brand et al. in prep.\\
TOOT16w & $ROz$ & $S_{1.4}>$30 mJy & $E\approx R<$19.5  & $B-R >$1.8 & 0.0151 &Brand et al. in prep.\\
TOOT08w & $ROz$ & $S_{1.4}>$30 mJy & $E\approx R<$19.5  & $B-R >$1.8 & 0.02786 &Brand et al. in prep.\\
TONS08q & $ROz$ & $S_{1.4}>$3 mJy  & $E\approx R<$19.83 & $B-R <$1.8 & 0.00688 &Brand et al. in prep.\\
Lacy    & $ROz$ & $S_{1.4}>$20 mJy &17$<E\approx R<$20.2& galaxies from APM&0.0122&\citet{lac}\\
TOOT    & $Rz$  & $S_{151}>$100 mJy& NONE               &NONE&& \citet{hr}\\
7CII    & $Rz$  & $S_{151}>$500 mJy& NONE               &NONE&& Willott et al. (2002)\\
Sadler  & $ROz$ & $S_{1.4}>$2.8 mJy&14$<B_j<$19.4& AGN$^{1}$ &0.099& Sadler et al. (2002)\\
2dF     & $Oz$  & NONE             & $B_j<$19.4 & NONE & 0.61 & Colless et al. (2001) \\
NVSS$^{2}$ & $R$& $S_{1.4}>$2.5 mJy & NONE & NONE &10.36 &Condon et al. (1998)\\
FIRST$^{3}$& $R$& $S_{1.4}>$2 mJy   & NONE & NONE & 2.61 &Becker et al. (1995)\\
Peacock \& Nicholson & $Rz$ & $S_{1.4}>$500 mJy & NONE$^{4}$ & NONE & 9.3 & \citet{pn}\\
\hline\hline
\end{tabular}
{\caption[Surveys]{\label{tab:samples} Table summarising the various samples used in and/or referred to in this paper. $R$ denotes radio galaxy surveys, $O$ denotes surveys with an optical magnitude cut and $z$ denotes spectroscopic redshift surveys. Notes: 1. AGN were identified using Principal Component Analysis (PCA) applied to the spectra \citep{fol}. 2. NVSS is 50 per cent complete at 2.5 mJy and 99 per cent complete at 3.5 mJy. 3. FIRST is 95 per cent complete at the given limiting flux densities. 4. \citet{pn} impose a redshift limit of $z$=0.1.   
}}
 \end{center}
 \end{table*}

\normalsize

The selection criteria for all TONS surveys were based on cross-matching positions of objects in radio and optical surveys. An initial selection was made of NVSS targets in the chosen area of sky with 1.4 GHz flux densities $S_{1.4}$ greater than 3 mJy (just above the completeness limit of $S_{1.4} \approx$ 2.5 mJy; \citealt{con}). Although the FIRST survey has better spatial resolution than NVSS, we used NVSS catalogue positions to do our matching. This is because FIRST is more likely to resolve out nearby, low surface brightness objects hence causing possible incompletenesses within the survey. A total of 1148 sources were selected in our survey area. 
The NVSS positions were matched with APM positions of objects with $E$-band magnitudes less than 19.5. The APM survey is a digitised sky survey of Palomar blue ($O$) and red ($E$) sky survey plates measured by the Automatic Plate Measuring (APM) machine in Cambridge \citep{mcm}. These magnitudes can be related to the more widely used Johnson magnitudes using:

\begin{equation}
E=R
\end{equation}
\begin{equation}
O=B-0.12(O-E)
\end{equation}

\noindent\citep{mcm}. 

\begin{figure*}
\begin{center}
\setlength{\unitlength}{1mm}
\begin{picture}(150,100)
\put(110,35){\includegraphics{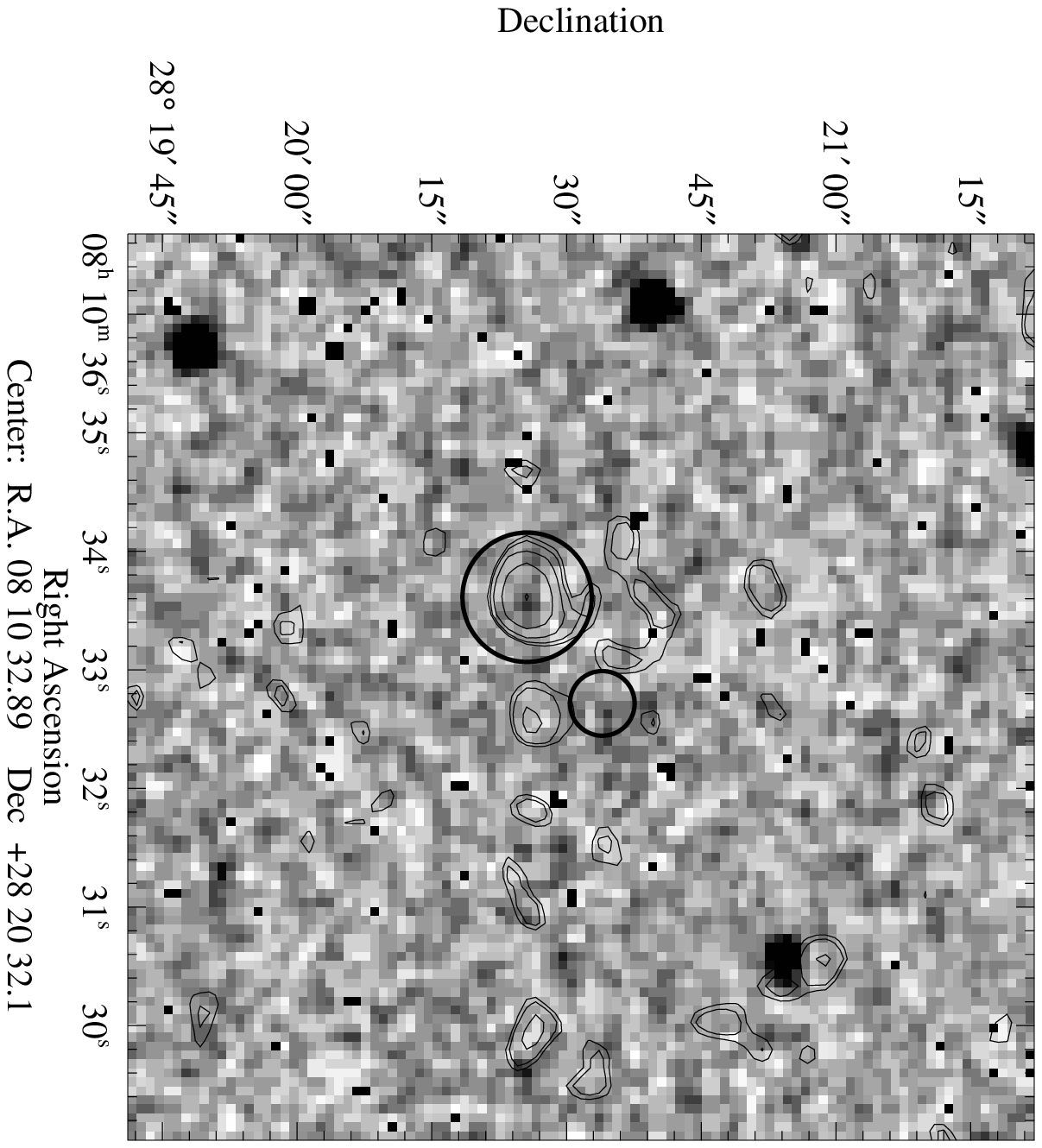}}
\put(170,35){\includegraphics{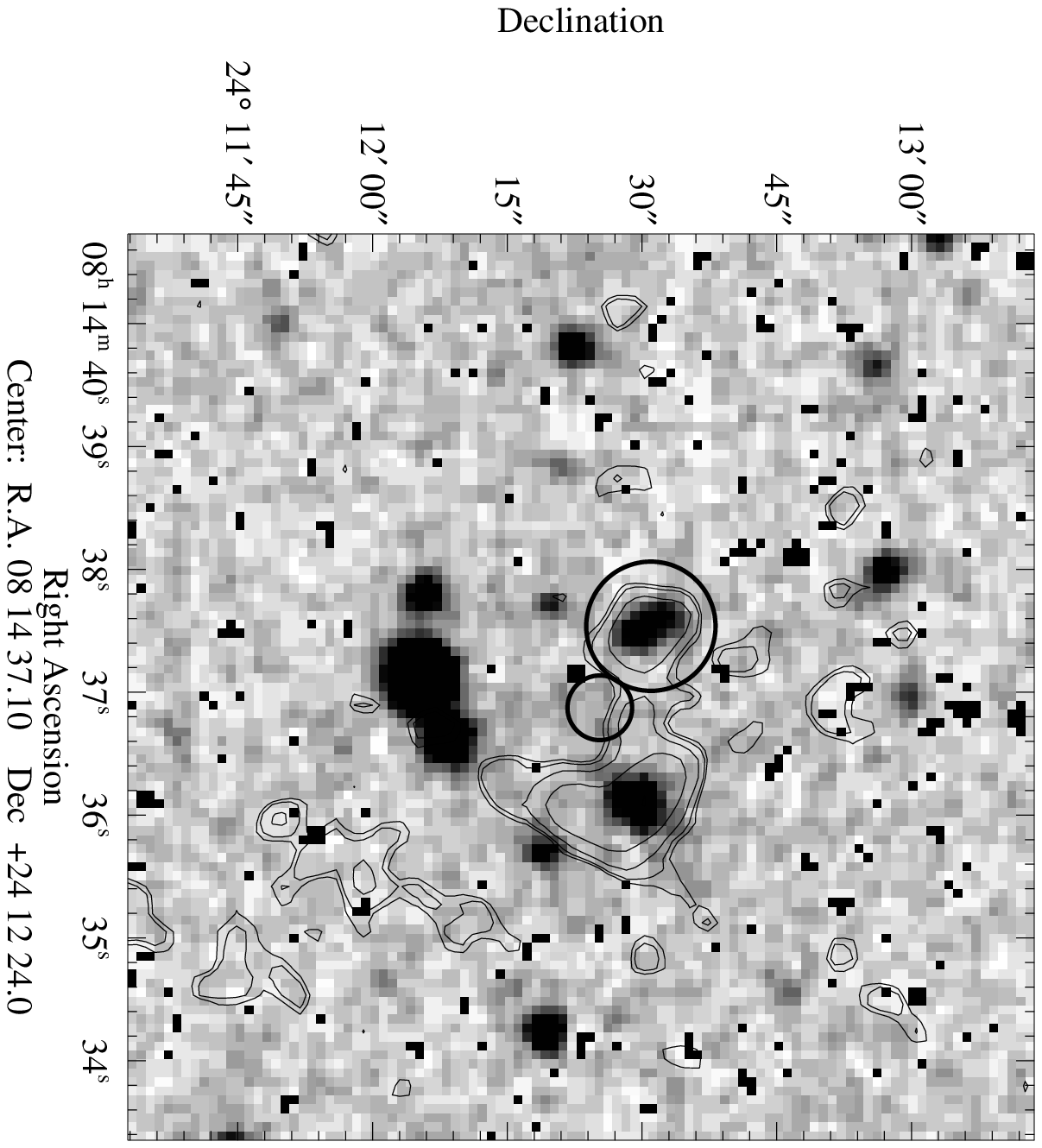}}
\put(110,-20){\includegraphics{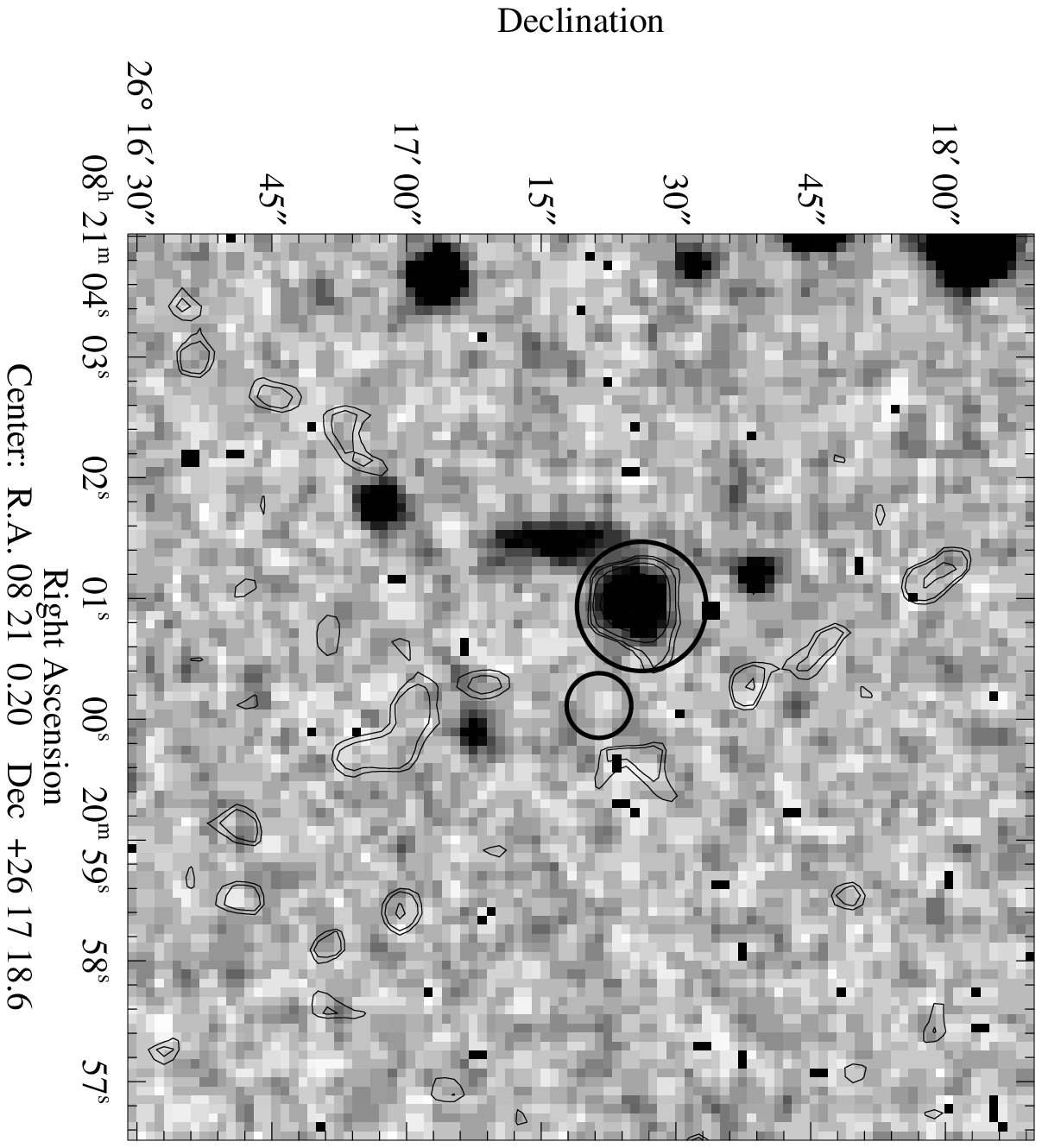}}
\put(170,-20){\includegraphics{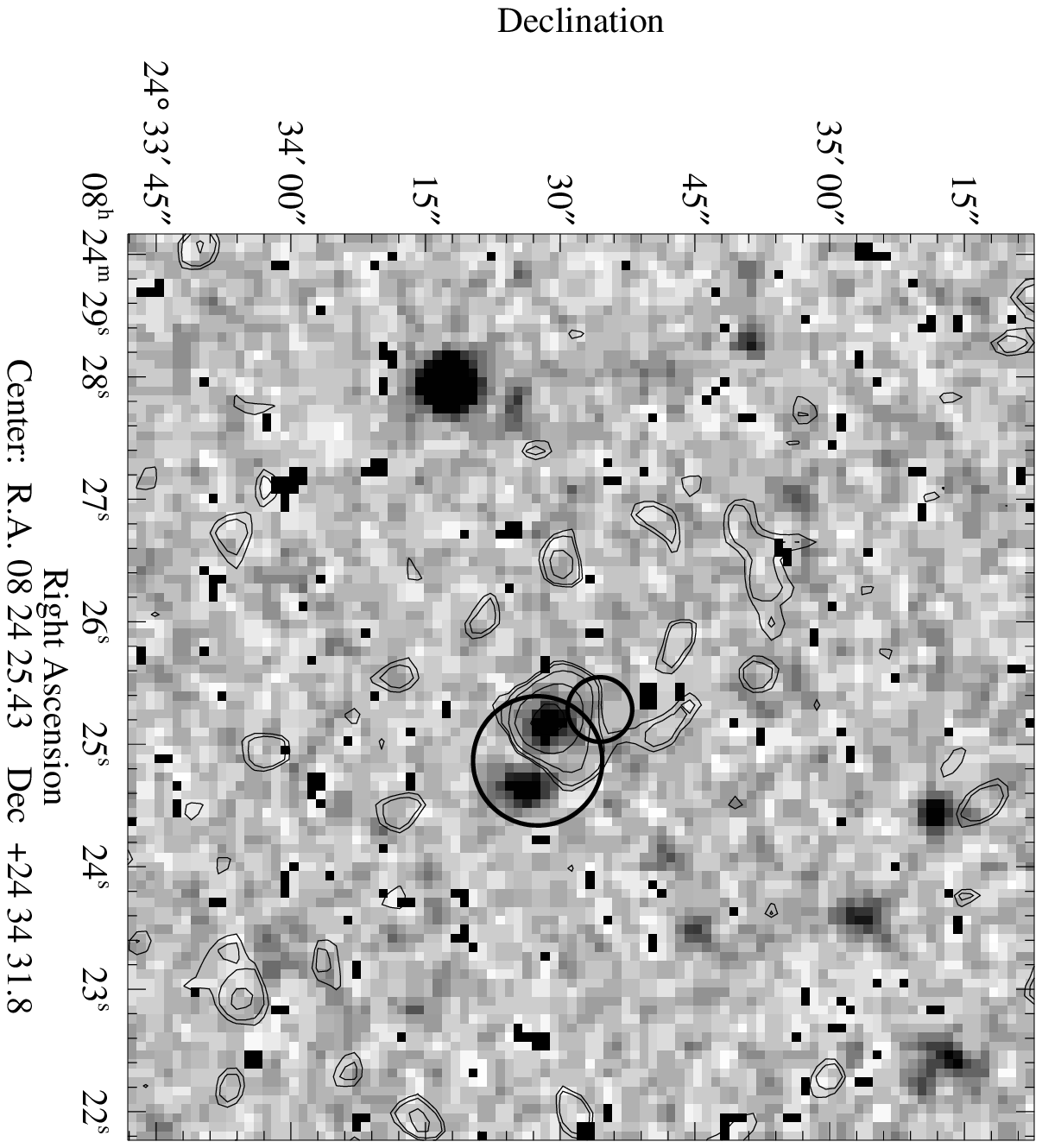}}
\end{picture}
\end{center}
{\caption[junk]{\label{fig:images} Four examples of radio galaxies selected in the TONS08 sample. The galaxies are (top left to bottom right): TONS08$\textunderscore$015,TONS08$\textunderscore$263,TONS08$\textunderscore$647 \& TONS08$\textunderscore$858. The small circle denotes the NVSS position and the large circle denotes the APM position. TONS08$\textunderscore$015 and TONS08$\textunderscore$647 show how the NVSS position can be substantially out if the radio flux density is small. The figure of TONS08$\textunderscore$263 shows how NVSS has treated what is actually two sources as a single extended source due to its lower resolving power. The figure of TONS08$\textunderscore$858 shows how the APM position can be wrong when it blends two nearby objects. 
}}
\end{figure*}

To ensure a complete, flux-density limited survey we adopted the following procedure. We selected any objects with APM and NVSS positions with an offset of $\le$ 20 arcsec from each other. We then plotted radio contours from the FIRST survey over optical POSS-II images and overplotted the positions of the APM and NVSS objects, identifying real identifications by eye. Although time consuming, it is important to do this as both the radio and the optical positions are subject to various complex uncertainties. Fig.~\ref{fig:images} shows some examples of objects where the offset between the APM and NVSS positions was greater than 10 arcsec. If we had selected radio galaxies by employing a simple cutoff offset between radio and optical positions we would have either missed real objects or would have contaminated our sample with miss-IDs.

The NVSS radio survey is thought to have rms positional uncertainties of $\ltsimeq$ 1 arcsec for relatively strong point sources ($S_{1.4} >$ 15 mJy) up to about 7 arcsec for the faintest detectable sources \citep{con}. TONS08$\textunderscore$015 and TONS08$\textunderscore$647 in Fig.~\ref{fig:images} show that in practise it can actually be more than this. The synchrotron emission detected in radio galaxies comes from the extended lobes associated with the jets. Fanaroff \& Riley class II (FRII) objects \citep{fr} often have two hot-spots where the jets terminate. Due to its lower resolution, the NVSS survey tends to find a position close to the centre of these double radio galaxies. If the lobes are asymmetric, the position can be less accurate and more worryingly, two close sources can be blended together and missed from the sample. TONS08$\textunderscore$263 in Fig.~\ref{fig:images} shows two radio sources with a very small angular separation. These have been mis-classified as the two lobes of an extended radio galaxy. It is especially important not to miss such close pairs of objects in clustering studies. As a further check, we obtained FIRST cutouts over a larger area for each NVSS position to check if two FIRST sources were blended into one NVSS source. This picked up two nearby sources which had already been picked up by eye and included in our sample. TONS08$\textunderscore$858 in Fig.~\ref{fig:images} shows an example of how the APM survey can blend two objects together. The survey has a scanning resolution of 1 arcsec.  In practise, objects too close together may be blended together making the positions less accurate. The APM catalogues are thought to be $\ge$99 per cent complete \citep{mcm}.
Although positions are accurate in the APM survey, the magnitudes can be out by 0.5 mag. The $O$ and $E$ magnitudes have been modified using corrections derived from comparison of the APM magnitudes to GSC-2 magnitudes for each POSS-II plate (R. White priv. com.). As our survey extends over two POSS-II plates, we modify the magnitude cut from our original target of 19.5 to 19.83 (the new limit of the most shallow plate). All following analysis uses these corrected values.

\begin{figure}
\begin{center}
\setlength{\unitlength}{1mm}
\begin{picture}(150,65)
\put(-2,-10){\includegraphics{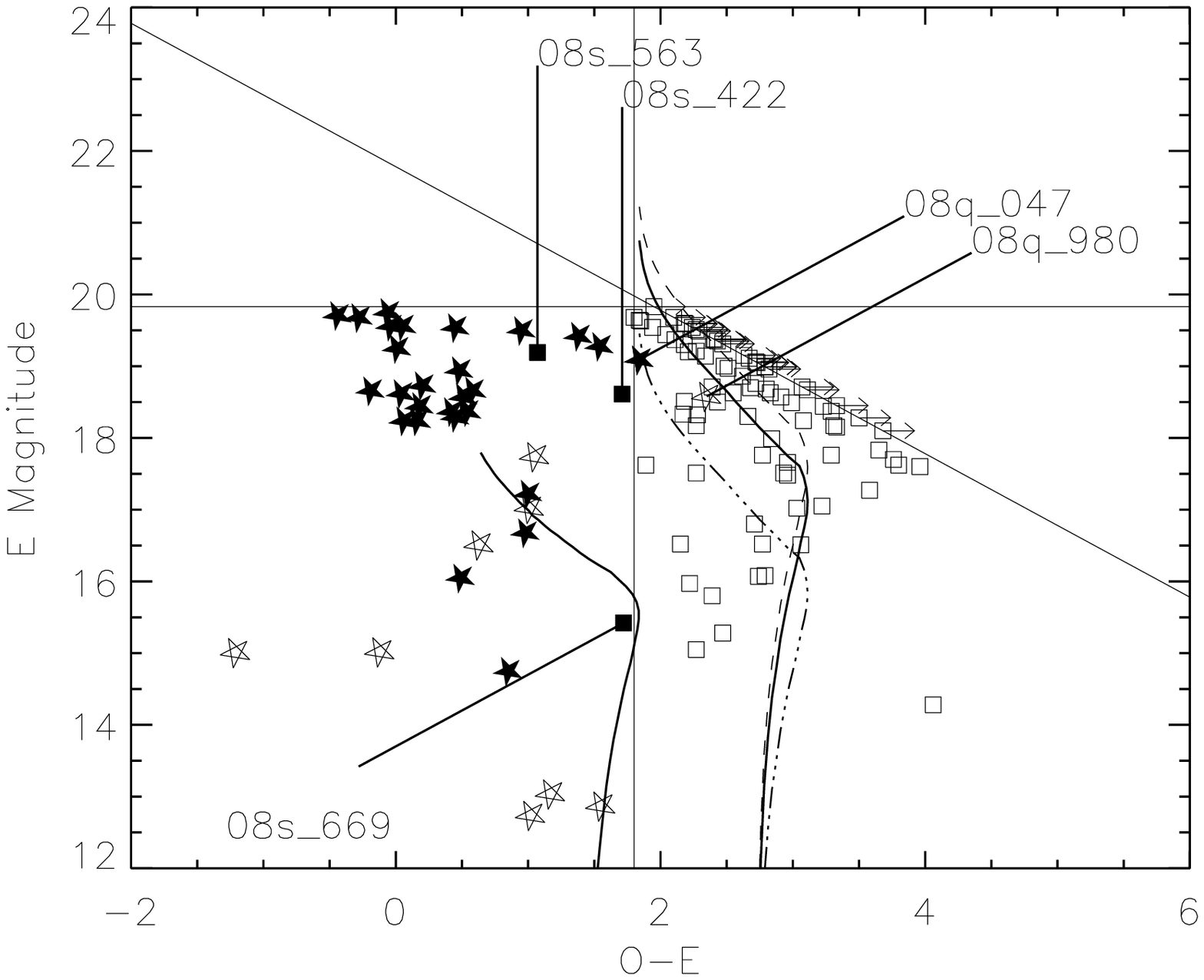}}
\end{picture}
\end{center}
{\caption[junk]{\label{fig:col_plot} The $O-E$ colour versus $E$ Magnitude plot for the TONS08 and TONS08q samples. TONS08 objects are squares, TONS08q objects are split into the lower redshift starburst galaxies (stars) and quasars (filled stars). The $O-E$=1.8 colour cut which divides the two samples is shown as well as the $E$=19.83 mag cut. Objects denoted by an arrow are such because APM $O$-magnitudes only go as faint as 21.48 and therefore only have their minimum colours shown. Overplotted are template SEDs from GISSEL (Bruzual \& Charlot 1993). To the right of the colour cut are the SEDs of an evolved stellar population. The solid line is for a galaxy with the average TONS08 absolute $R(\approx E)$ magnitude (-22.89). The dashed and dash-dot-dotted lines are $L\star$ and $5L\star$ respectively. To the left of the colour cut is an SED of a young stellar population with the average TONS08q star forming population absolute $R(\approx E)$ magnitude (-23.94). To see the redshift corresponding to each $E(\approx R)$ magnitude, see Fig.~\ref{fig:rz}.}}
\end{figure}

The final selection criteria was that $O-E\ge$1.8. This can be related to the more widely used Johnson filter colour, $B-R$ by:

\begin{equation} 
B-R=0.88(O-E)
\end{equation}

\noindent\citep{mcm}. This cuts out all the bluer objects which tend to be stars (which by coincidence lie close to the line of sight to the radio source), quasars (which generally have much higher redshifts than that under interest) and starburst galaxies (which tend to be at low redshifts due to their low radio luminosity).

Fig.~\ref{fig:col_plot} shows the loci of AGN radio galaxies, starburst galaxies and quasars in the optical magnitude versus colour plane. The template spectral energy distributions (SEDs) were obtained from the Galaxy Isochrone Synthesis Spectral Evolution Library (GISSEL; \citealt{bc}). The SEDs show the paths of galaxies with different absolute magnitudes. For the evolved ellipticals, we use a B magnitude value of $M_{*}$=-20.9 for $L\star$ \citep{hl}. The average absolute R magnitude of the TONS08 sample is $M_{*}$=-22.89. This corresponds to $\approx 1.5L\star$ and is shown by the solid line. 

It can be seen that the colour cut will divide the populations fairly cleanly. To check that no AGN type galaxies were missed by the colour cut (either unusually blue galaxies or galaxies with incorrect colours on APM plates) we also obtained spectra for all objects with $O-E\le$1.8 (the TONS08q sample; Brand et al. in prep.). Three objects with $O-E\le$1.8 were moved to the TONS08 sample (TONS08$\textunderscore$422,TONS08$\textunderscore$563 and TONS08$\textunderscore$669). The spectra of TONS08$\textunderscore$422 and TONS08$\textunderscore$563 show clear characteristics of an old elliptical type population as does TONS08$\textunderscore$669 which is in the 7CRS \citep{wil02}. Two objects with $O-E\ge$1.8 were rejected from the TONS08 sample and moved into the TONS08q sample (TONS08$\textunderscore$980 and TONS08$\textunderscore$047). TONS08$\textunderscore$980 is identified as a nearby star-forming galaxy: it has strong S [II] and [O II] emission lines and is at very low redshift. TONS08$\textunderscore$047 is in the TOOT08 survey and is identified as a high redshift quasar. These are annotated in Fig.~\ref{fig:col_plot}.

The final number of objects in the TONS08 sample is 84. Information about these sources is shown in Table.~\ref{tab:summary}. Four of the five 7C-II objects defining the 7C-II redshift spike discussed by \citet{rhw} have also been selected in the TONS08 sample; 7C0818+2932, very unusually, is too optically faint \citep{wil02}. Of the two additional objects, 7C0811+2838 is selected (TONS08$\textunderscore$279) and NV0840+2805 falls outside the survey limits (but is picked up by the wider TONS08w survey). The 7C-II objects which match the TONS08 super-structure members are shown in Table.~\ref{tab:summary}.
The TONS08q sample comprises 40 objects. and will be the subject of a forthcoming paper.

\section{OBSERVATIONS}
\label{sec:spectra}
\subsection{Optical spectra}

Optical spectra were obtained during the period October 2000 - February 2002 on the 2.6m Nordic Optical Telescope (NOT) using the Andalucia faint object spectrograph, the 4.2m William Herschel telescope (WHT) using ISIS, the 2.7m Smith reflector at McDonald with the Imaging grism instrument (IGI) \citep{hill02}, and the Hobby-Eberly telescope (HET) using the Marcario low resolution spectrograph (LRS) \citep{hill98}. The IGI has recently been upgraded.  All observations from January 2002 were with the new volume phase holographic grism. 

Most objects in the TONS08 survey were identified as moderate redshift radio galaxies as expected. Spectra for the complete sample can be found in the complete paper at: 

\noindent http://www-astro.physics.ox.ac.uk/$\sim$brand/08$\textunderscore$paper.ps.gz. 

\begin{figure*}
\begin{center}
\setlength{\unitlength}{1mm}
\begin{picture}(200,80)
\put(155,-5){\includegraphics{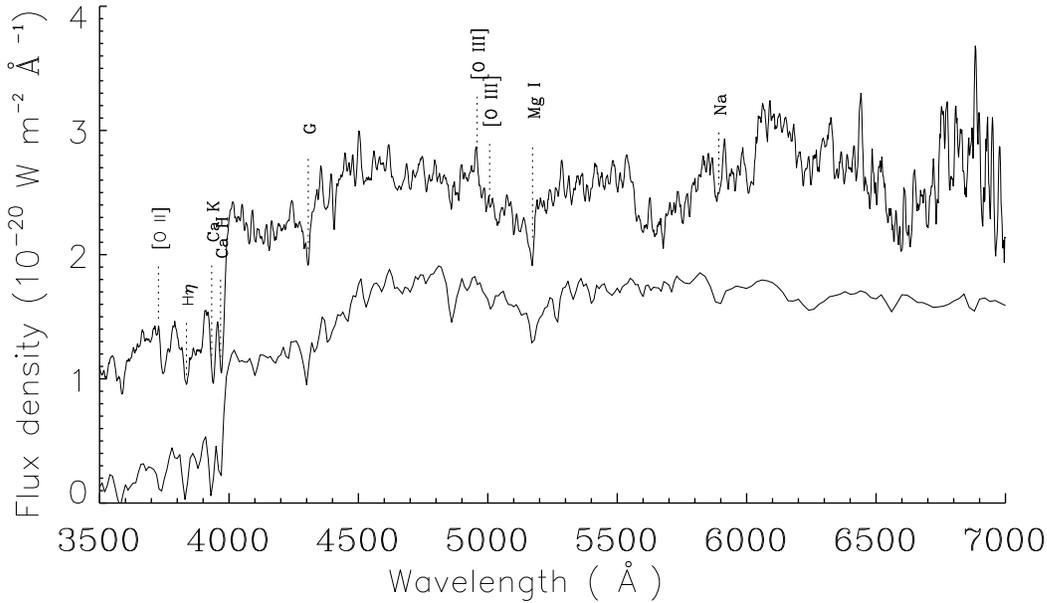}}
\end{picture}
\end{center}
{\caption[junk]{\label{fig:template} The de-redshifted composite spectrum for the best 50 spectra in the TONS08 sample. The main characteristic features are the $4000 ~ \rm \AA$ break and the nearby Ca K and Ca H lines. Other dominant features are the G and Mg I absorption lines. Also shown is a \citep{bc} model stellar population which formed in an instantaneous burst 2.0 Gyr ago (displaced by -1x$10^{-20}\rm{Wm}^{-2}\rm{\AA}^{-1}$ for clarity). Note the lack of strong emission lines in this composite spectrum. 
}}
\end{figure*}

\noindent They are sorted by redshift for clarity. In most cases, redshifts were determined from absorption lines (very few objects exhibited emission lines). The estimated redshifts are presented in Table.~\ref{tab:summary}. These were checked by performing a cross correlation between the spectra and both a rest frame composite spectrum obtained from the data (see Sec.~\ref{sec:spectra} and Fig.~\ref{fig:template}) and an evolved population GISSEL model \citep{bc}.  

\begin{figure}
\begin{center}
\setlength{\unitlength}{1mm}
\begin{picture}(150,65)
\put(-2,-10){\includegraphics{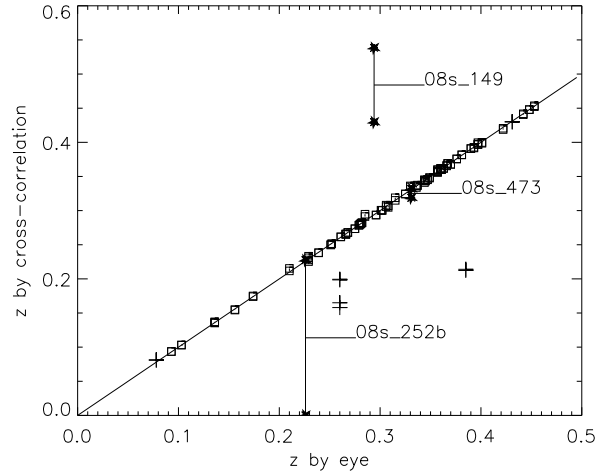}}
\end{picture}
\end{center}
{\caption[junk]{\label{fig:zcor} The by-eye redshift versus both cross-correlated redshifts for the TONS08 sample. The three objects with the most unclear redshifts are labelled and denoted as stars, objects with strong lines are denoted as plus signs and all other objects are denoted as squares.
}}
\end{figure}

Fig.~\ref{fig:zcor} shows the redshift obtained by eye against the averaged cross-correlated redshift for the TONS08 survey. There is extremely good agreement between the two different methods of obtaining redshifts for the vast majority of objects. Six objects have strong line emission and although lines were taken out for the cross correlation, the redshifts obtained by cross-correlation showed poor agreement with the redshifts by eye. This is not surprising as these objects will be of different spectral type and observations would have been cut short once it was clear that a redshift could be obtained. In all future analysis, we take the redshift obtained by eye for these objects. 

There are only three objects for which we don't have an unequivocal redshift estimate. TONS08$\textunderscore$252B has a very poor quality spectrum and only one of the cross-correlation redshift estimations matched that obtained by eye. However, as this object is very close on the sky to TONS08$\textunderscore$252A, it is very probable that the redshifts will be similar; we take the redshift estimate to be real. TONS08$\textunderscore$149 also has a poor quality spectrum. The by-eye redshift estimate does not agree with those obtained by cross-correlation. These redshift estimates are probably the result of flux calibration errors which cause a jump in the spectrum where the red and blue arms of the WHT ISIS are combined. TONS08$\textunderscore$473 possibly has a lower redshift. However, the cross-correlation estimates agree with the higher estimate so we shall assume that this is correct. We note here that none of the three objects are in either of the super-structures. Consequently, if any of these redshifts are incorrect, it could only boost the significance of the super-structures. 

\begin{figure}
\begin{center}
\setlength{\unitlength}{1mm}
\begin{picture}(150,65)
\put(-2,-10){\includegraphics{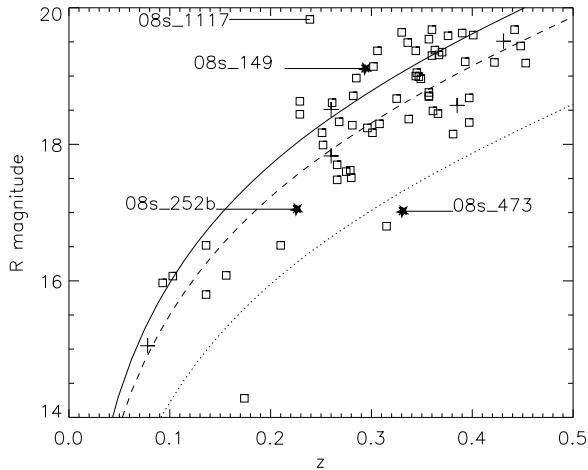}}
\end{picture}
\end{center}
{\caption[junk]{\label{fig:rz} The $R$ magnitude versus redshift plot for the TONS08 sample. The three objects with the most unclear redshifts are labelled and denoted as stars, objects with strong lines are denoted as plus signs and all other objects are denoted as squares. Overplotted are template SEDs of evolved stellar populations from GISSEL \citep{bc}. The dashed line is for a galaxy with the average TONS08 absolute $R$ magnitude (-22.89). The solid and dotted lines are $L\star$ and $5L\star$ respectively. Note that the R magnitudes are only accurate to $\approx$0.3 mag (see Sec.~\ref{sec:model_z}). TONS08\textunderscore1117 is marked as it appears to be unusually faint for its redshift. This source is discussed in Sec.~\ref{sec:spectra}.
}}
\end{figure}

We plot the $R$ magnitudes against (by eye) redshifts for the TONS08 sample in Fig.~\ref{fig:rz}. The three objects with the most inconclusive redshifts lie in approximately the same distribution as all other objects in the survey. Also, the redshifts approximately follow the expected underlying redshift distribution. Therefore, we take these redshifts to be correct in all following analysis. All following analysis is done on the redshifts obtained by eye. The only source with a noticeably faint $R$ magnitude for its redshift, is TONS08\textunderscore1117 (see Fig.~\ref{fig:rz}). This redshift is fairly secure, with the two methods of cross-correlation agreeing with the by eye redshift (Table.~\ref{tab:summary}). Assuming that the APM $R$ magnitude is correct, this may be an unusual source.
 
A composite spectrum is shown in Fig.~\ref{fig:template}. To construct the composite spectrum, the 50 best TONS08 spectra were trimmed and shifted to their rest-frame. The spectra were then normalised using their integrated flux over the range 4000-4500$ \rm \AA$ and then combined by taking their median value. We have compared this spectrum to a \citet{bc} model stellar population which formed in an instantaneous burst 2.0 Gyr ago. The good agreement shows these objects are consistent with an old, passively evolving stellar population. 
The prominent features used to identify the redshifts \citep{san} of the radio galaxies are over-plotted. 

The majority of the TONS08q sample were low radio flux density ($S_{1.4}\le10$ mJy), low redshift ($z\le$0.1) starburst galaxies or higher flux density, high redshift quasars (see Fig.~\ref{fig:col_plot}). Only one object in TONS08q falls between $z$=0.11 and $z$=0.69. Interestingly this object is at $z$=0.283, at the first redshift spike in the TONS08 sample. This object is an X-ray selected QSO \citep{wei}. Although interesting, we will not include this in any further analysis here.

\subsection{The 3-D distribution}

\begin{figure}
\begin{center}
\setlength{\unitlength}{1mm}
\begin{picture}(80,115)
\put(-5,120){\includegraphics{08s_3d_redshifts_bw.ps}}
\put(0,70){\includegraphics{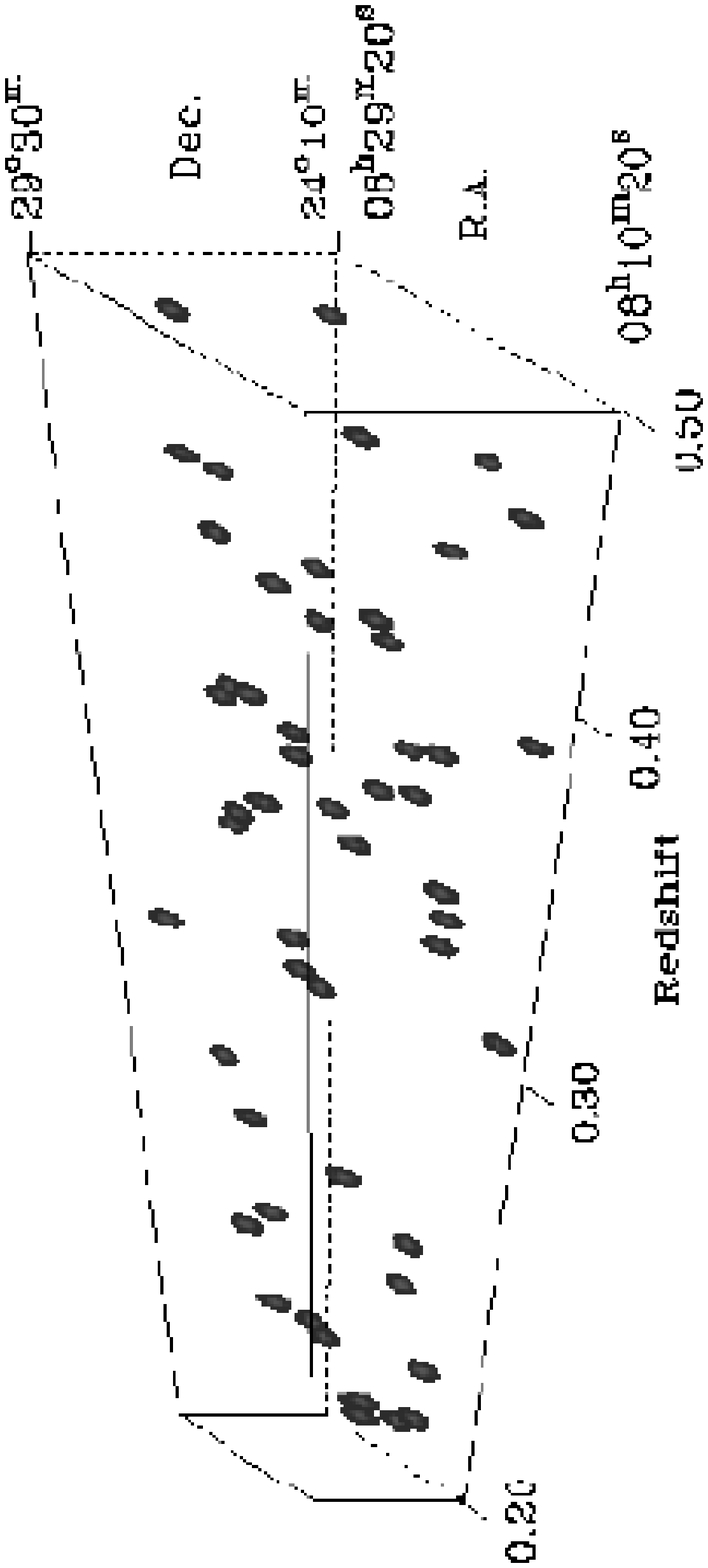}}
\put(0,40){\includegraphics{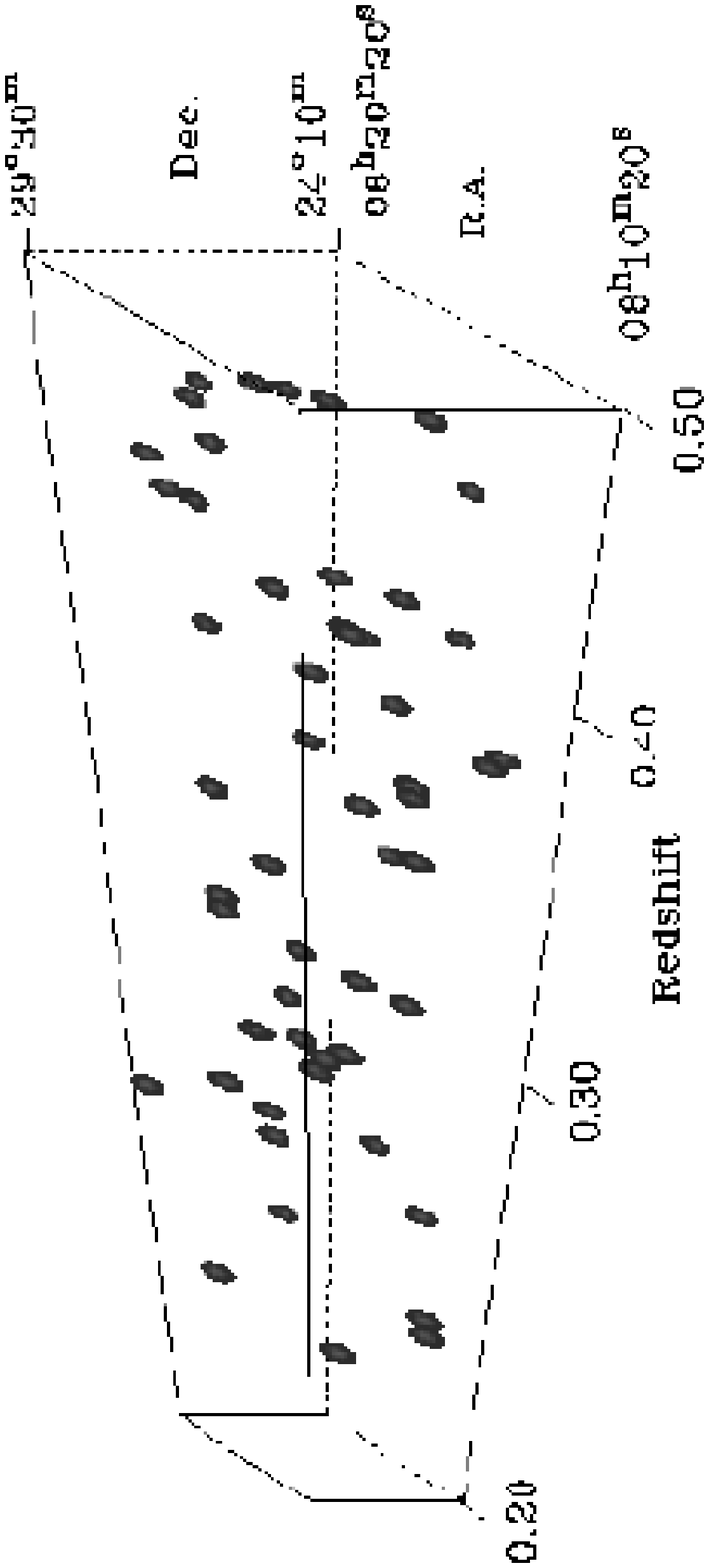}}
\end{picture}
{\caption[3-D distribution of TONS08 sample]{\label{fig:3d} Representation of the three-dimensional distribution of the TONS08 sample between $z$=0.2 and $z$=0.5 (top). Note the effect of the selection function in redshift as plotted in Fig.~\ref{fig:zdistn_rlf}. Clear groupings of radio galaxies can be seen at $z\approx$0.27 and $z\approx$0.35. Also plotted below are two random samples with a selection function from the model redshift distribution (Fig.~\ref{fig:zdistn_rlf}). TONS08 looks significantly more clustered than either of the randomly distributed samples.}} 
\end{center}
\end{figure}

A representation of the three-dimensional distribution of the TONS08 sample is shown in Fig.~\ref{fig:3d}. Clustering of objects can be clearly seen at a redshift of $z$ $\approx$ 0.27. There are at least 13 objects tracing a super-structure with a spatial extent of at least 80 $\times$ 100 $\times$ 100 $\rm {Mpc}^3$ (the sense is $\Delta \rm{RA} \times$ $\Delta \rm{DEC} \times$ $\Delta z$). The size may be limited by the spatial extent of our sky coverage in RA and DEC. There is also a second peak in the distribution at $z\approx$0.35. This structure is approximately 310 Mpc away from the lower redshift structure (from centre to centre measured in co-moving units) and at least 12 objects trace a super-structure with a spatial extent of at least 100 $\times$ 100 $\times$ 100 $\rm {Mpc}^3$ (again limited in RA and DEC by the sky coverage).

\begin{figure}
\begin{center}
\setlength{\unitlength}{1mm}
\begin{picture}(150,65)
\put(-2,-10){\includegraphics{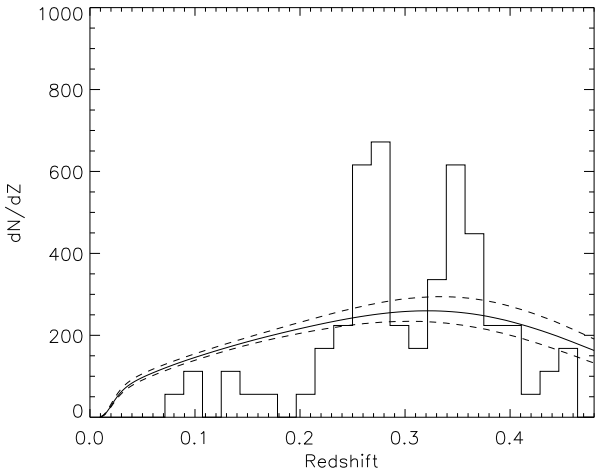}}
\end{picture}
\end{center}
{\caption[junk]{\label{fig:zdistn_rlf} The redshift distribution of the TONS08 sample with the model redshift distribution overplotted (solid line). The $\pm 1\sigma$ errors on the model are overplotted (dashed lines). We plot the distribution using 28 redshift bins between $z$=0 and $z$=0.5.
}}
\end{figure}

\section{ANALYSIS}

\subsection{A model redshift distribution}
\label{sec:model_z}

The redshift distribution of the TONS08 sample is shown in Fig.~\ref{fig:zdistn_rlf}. Overplotted is a model distribution normalised to the same sky area and evaluated for the same optical and radio selection criteria as TONS08. 

We obtained the model redshift distribution by integrating a fitted model distribution function of radio galaxies. This was achieved by using the maximum likelihood method of \citet{mar} to estimate the best fit parameters for the bivariate (radio and optical) luminosity function (BLF). We required a data set over a sufficiently large area to reduce the effects of cosmic variance due to large-scale structure. The data used to constrain this model were from \citet{sad}. This cross matches NVSS sources with the first 210 fields observed in the 2dFGRS and covers an effective area of 325 deg$^2$ (13 times larger than TONS08). There are two distinct populations in this sample; AGN (60 per cent) and star-forming galaxies (40 per cent). We used only the AGN as the colour cut in TONS08 cuts out the latter population.
The BLF was modelled with the following function:
 
\begin{equation} 
\rho\left(L,M_{b},z\right)=A\left(1+z\right)^{q}\left(\frac{L}{L_r}\right)^{-\gamma}e^{-\frac{\left(M_{b}-<M_{b}>\right)^{2}}{2\sigma^{2}}}
\end{equation}
 
\noindent where $\rho\left(L,M_{R},z\right)$ is the source number per unit co-moving volume per unit log$_{10}L$ per unit $M_{b}$ as a function of radio luminosity $L$, optical absolute $b_j$ magnitude $M_{b}$ and redshift, $z$. $A$, $q$, $\gamma$ and $\sigma$ were the parameters to be fitted. The best fit values for the parameters are as follows: 
\noindent log$_{10}A$=-28.851, log$_{10}\sigma$=-0.096, $\gamma$=1.710, $q$=1.876.

The redshift distribution was obtained by integrating the BLF over the selection criteria using a k-correction obtained from template SEDs \citep{bc}. 

Errors on the distribution were found by performing a Monte Carlo simulation. As it is likely that other errors will also contribute including incompletenesses in both the radio and optical surveys at faint flux densities, we included these as follows. The NVSS catalogue is $\approx$ 90 per cent complete at S$_{1.4}$=3 mJy \citep{con}. There is also a 2 per cent systematic surface density fluctuation across the sky \citep{bw}. This will be a negligible effect compared to completeness problems at the magnitude limit of the APM (and hence TONS08) survey. Before correcting for plate-to-plate variations, the APM magnitudes have a global rms uncertainty of 0.5 mag \citep{mcm}. This will be reduced by the plate-to-plate corrections. We assume an error of 0.3 mag. This is the difference between the magnitude corrections of the two POSS-II plates in the TONS08 region. We incorporated this into the error on the redshift distribution by determining the R magnitude limit as a Gaussian distributed value with mean 19.83 and standard deviation 0.3. The Monte Carlo method also allowed for errors in the derived model parameters.   

The redshift distribution was then calculated for parameter combinations whose maximum likelihood values fall within $\pm$1$\sigma$. This model is the subject of a forthcoming paper (Brand et al. in prep.). 

\begin{figure}
\begin{center}
\setlength{\unitlength}{1mm}
\begin{picture}(150,65)
\put(-2,-10){\includegraphics{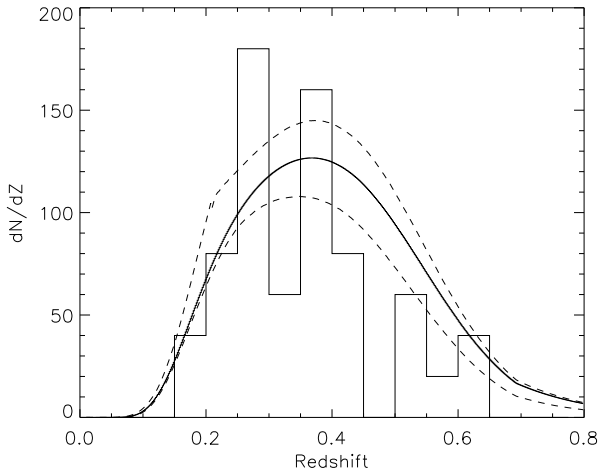}}
\end{picture}
\end{center}
{\caption[junk]{\label{fig:mark} The redshift distribution of the \citet{lac} sample with the model redshift distribution overplotted. The $\pm$1$\sigma$ errors on the model are overplotted (dashed lines). We plot the distribution for 16 redshift bins between $z$=0 and $z$=0.8.
}}
\end{figure}

The 2dFGRS has a median redshift of 0.1 \citep{fol}. As a consequence, our redshift distribution will be less well constrained at redshifts greater than $\approx$ 0.3. As a check, we obtained a model redshift distribution for the sample obtained by \citet{lac}. Fig.~\ref{fig:mark} shows that the model provides a good fit to his data, even out to high redshift. It also shows two redshift spikes in the data although these do not appear to be very significant. We consider this sample further in Sec.~\ref{sec:lacy}. 

A comparison of the TONS08 and the model redshift distribution in Fig.~\ref{fig:zdistn_rlf} shows that large-scale structure has changed the redshift distribution of our sample significantly from that expected for a typical area of sky. The model predicts 95$\pm11$ radio galaxies in this area of sky. Although this is more than the 84 observed sources, the difference is not significant. \citet{ben} have already shown that large-scale structure can produce $\sim 5$ per cent variations in the source counts over $5\times 5$ deg$^2$ regions. The signatures of large-scale structure in TONS08 seem clear: two prominent spikes at $z\approx$0.27 and $z\approx$0.35 with, perhaps, void-like regions at lower redshifts, accounting for the low overall number of radio galaxies. 

\subsection{The significance of clustering in redshift space}
\label{sec:sig}

More than one-half of the total sample lie in the regions $z$=0.24-0.28 and $z$=0.33-0.38. This is extremely unlikely to arise from Poisson fluctuations associated with sampling the expected redshift distribution (Fig.~\ref{fig:zdistn_rlf}). 

\begin{figure*}
\begin{center}
\setlength{\unitlength}{1mm}
\begin{picture}(80,65)
\put(-50,-10){\includegraphics{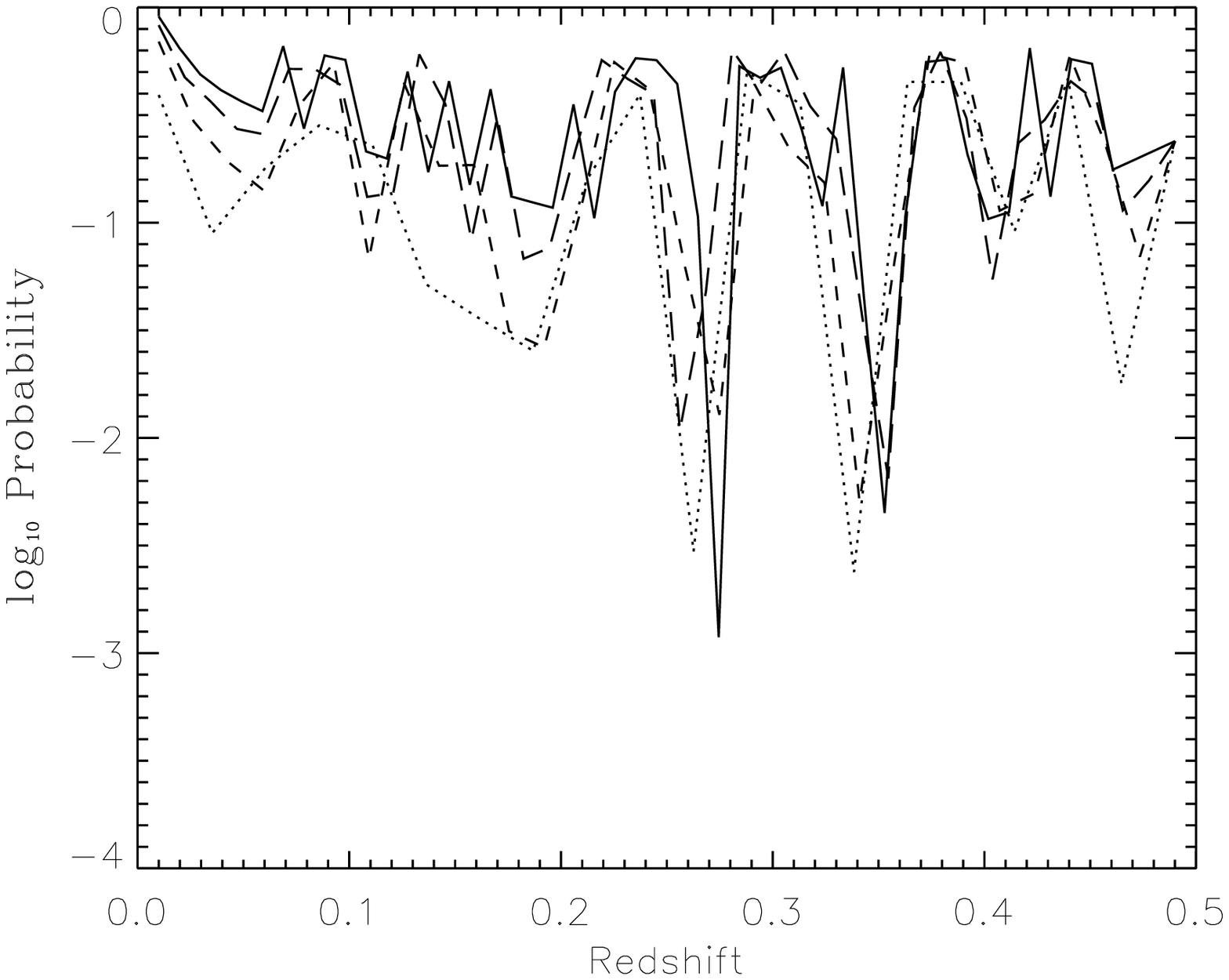}}
\put(30,-10){\includegraphics{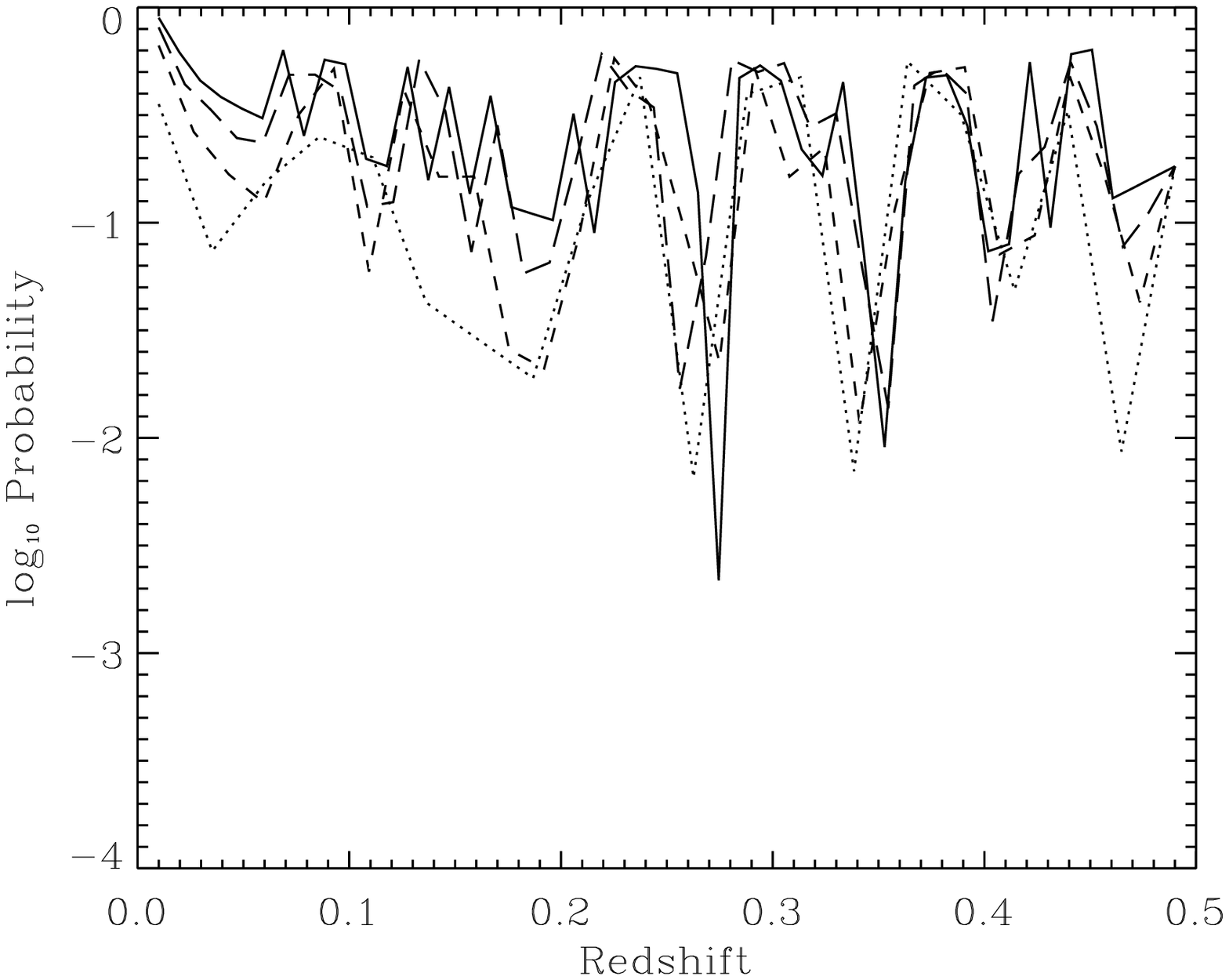}}
\end{picture}
{\caption[simp_prob]{\label{fig:simp_prob} The probability that the number of galaxies in each redshift bin could be greater than the actual number given the number predicted by the model (if the number of galaxies is greater than that predicted by the model) or the probability that the number of galaxies in each redshift bin could be less than the actual number given the number predicted by the model (if the number of galaxies is less than that predicted by the model). The results are plotted for 50 bins (solid line), 40 bins (long dashed line), 30 bins (dashed line) and 20 bins (dotted line). The figure on the left shows the results with the four 7CRS galaxies at $z\approx0.27$ omitted from the analysis. The figure on the right shows the results calculated for the same data but for the model redshift distribution at the $+ 1 \sigma$ level from Fig.~\ref{fig:zdistn_rlf}. 
}}
\end{center}
\end{figure*}

We first explored the significance of clustering with a simple method using Poisson statistics in each redshift bin to calculate the probability that the number of galaxies could be greater than the actual number, given the number predicted by the model (if the number of galaxies is greater than that predicted by the model) or the probability that the number of galaxies could be less than the actual number given the number predicted by the model (if the number of galaxies is less than that predicted by the model). We did this analysis excluding the 4 7C-II objects at redshift $z\approx$ 0.27 in order to seek evidence for redshift spikes independent of that seen in the 7CRS \citep{rhw}. Fig.~\ref{fig:simp_prob} shows this probability plotted for different binning intervals. It can be seen that different binning intervals make little difference to the final result. For 20 bins, the probability in the redshift interval $z$=0.34 to $z$=0.36 is 0.002. Because there are 20 independent bins, one expects such a low value in 4 per cent of a large set of realisations. In the redshift interval $z$=0.26 to $z$=0.28, the probability is 0.003. If this was at a random redshift, we would expect a redshift peak of this size or greater in 6 per cent of random realisations. However, because this redshift peak occurs in the same place as independent data from 7C-II \citep{rhw} and the TONS08 survey was designed to find an independent signal in this particular bin, we would expect a peak at this same redshift in only 0.3 per cent of random realisations. There is also a suggestion of a void region between redshifts 0.1 and 0.2. Although in void regions the small numbers involved prevent this method from giving meaningful statistics, a void region within the survey would not be unexpected given the number of voids in large surveys such as the 2dFGRS and the filamentary distribution predicted by N-body simulations \citep{jen}.

We performed this analysis for redshift distributions corresponding to the $+ 1 \sigma$ error on the model redshift distribution in Fig.~\ref{fig:zdistn_rlf}. For 20 bins, the probability in the redshift interval $z$=0.34 to $z$=0.36 increases to 0.007. One expects such a low value in 14 per cent of a large set of realisations. In the redshift interval $z$=0.26 to $z$=0.28, the probability increases to 0.006. We would expect a peak at this same redshift in only 0.6 per cent of random realisations (given again that we are searching for a signal in a specific bin). We also note that our error analysis does not allow fully for inadequacies in the parameterised model. Specifically, the true uncertainty in the normalisation at 0.3$\le z \le$0.5 (i.e. beyond the redshift at which the model is directly constrained by the 2dFGRS data) may well be larger than appears to be the case from Fig.~\ref{fig:zdistn_rlf}. For these reasons, we feel that evidence for the $z$=0.35 super-structure is somewhat less secure the super-structure at $z$=0.27.

As a further check, we followed the method of \citet{ste} to estimate the significance of any number counts above Poisson expectations. \citet{ste} smooth their redshift distribution but we prefer to use our model distribution. We again took out the 7CRS radio galaxies to check for independent evidence.
 We considered each $N$ neighbouring galaxies and calculated whether their redshifts were closer together than expected from the Erlangian distribution. In order to implement this method, we transformed all redshifts into a coordinate system in which the redshift distribution is flat. We could then search for significant clustering above this flat background. We transformed our redshifts into a coordinate system $t$ using

\begin{equation}
t(z)=\int_{0}^{z}P(z) \ \rm{d}z \ ,
\end{equation}

\noindent where $P(z)$ is the model redshift distribution normalised so that $\int P(z)\rm \ {d}z=1$.

We then calculated the probability that the galaxies would be contained in an interval smaller than the observed interval $t2$

\begin{equation}
\zeta=p(t\le t2)=\rm{I}_{x}(N1+N3-1,N2-1) ,
\end{equation}

\noindent where $N2$ is the number of galaxies in interval $t2$, $N1$ is the number of galaxies with redshifts less than the super-structure member with the lowest redshift (in the interval $t1$), $N3$ is the number of galaxies with redshifts greater than the
super-structure member with the highest redshift (in the interval $t3$), $I_{x}$ is the incomplete beta function \citep{pre} and $x=(t1+t2)/(t1+t2+t3)$.

We calculated $\zeta$ for each set of $N2$ of neighbouring galaxies where $N2=$2,3,4.... To find the significance of each candidate super-structure, we created a set of simulated samples which had redshifts drawn from the same model redshift distribution. We then compared the $\zeta$ of the candidate super-structure to that of the simulated data. If the candidate $\zeta$ was smaller than only 1 in 100 simulated datasets then we assigned that super-structure a significance of 99 per cent.

We found that the two super-structures are both significant at the $\ge$99 per cent level. The intervals that maximise the significance of the redshift peaks are $z$=0.233-0.285 (27 radio galaxies) and $z$=0.33-0.367 (23 radio galaxies). 

\subsection{Calculating the radio galaxy overdensity}

Because TONS08 is only $\approx$80 Mpc wide (at redshift 0.27), there is an obvious danger of introducing bias by using the redshift intervals which maximise the significance of the redshift peaks (which correspond to co-moving depths of 195 Mpc and 135 Mpc) to estimate the radio galaxy overdensity. 
To match the data to theoretical overdensities, we chose instead to calculate overdensities in 50 Mpc radius spheres centred at the peaks of the redshift spikes. In this case we find 13 and 12 radio galaxies in the $z$=0.27 and the $z$=0.35 super-structures respectively. Of the 5 7CRS radio galaxies in the larger sample defined by $z$=0.233-0.285, only two (TONS08$\textunderscore$526 and TONS08$\textunderscore$669) are in the $z$=0.27 super-structure as defined by a sphere of radius 50 Mpc. We note that we have taken the centre of the TONS08 field for the centre of the 50 Mpc radius sphere. If we choose to maximise the number of radio galaxies in a 50 Mpc radius sphere, we could move the angular position of the centre of the sphere until we found the maximum number possible. In this case we find 15 radio galaxies in both the $z$=0.27 and $z$=0.35 super-structures. This would give the super-structures higher overdensities. 

Note also that we haven't taken into account redshift space distortions. If the super-structure is collapsing, this will mean there are actually fewer radio galaxies within a 50 Mpc radius sphere. The effects of this are discussed in more detail in Sec.~\ref{sec:discussion}.

We calculate the radio galaxy overdensity using 

\begin{equation}
\delta_{\rm{gal}}=\frac{N}{N_{\rm{model}}}-1
\end{equation}

\noindent where $N$ is the number of radio galaxies in the super-structure and $N_{\rm{model}}$ is the number predicted by the model. We determined $N_{\rm{model}}$ using a Monte-Carlo method. For each realisation, we distributed 84 objects at a random RA and DEC within the survey limits and at a random redshift weighted by the model redshift distribution. We then counted the number of these objects 50 Mpc away from the centre of each super-structure and averaged over the number of realisations. We predict there should be an average of 4.04 and 2.72 radio galaxies in the volume of the $z$=0.27 and $z$=0.35 super-structures respectively. The number predicted in the $z$=0.27 super-structure volume is more than that of the $z$=0.35 super-structure even though the model redshift distribution is higher at $z$=0.35. This is because the 50 Mpc sphere takes up less of the survey area at high redshift (see Fig.~\ref{fig:2d}).

We assume a Poisson error on the number of radio galaxies $N$ found within the 50 Mpc sphere. We also calculate the probability distribution $P(N_{\rm{model}})$of the expected number of radio galaxies $N_{\rm{model}}$ from the errors on the model redshift distribution. We determine the joint probability of finding this overdensity $OD$ given the data (Poisson and model distributions) using Bayes theorem (e.g. \citealt{siv}):

\begin{equation}
P(OD|data) \propto P(data|OD) \times P(OD). 
\end{equation}

\noindent P(OD) is simply a prior which we set to 1 if $\delta_{\rm{gal}}>-1$ and zero otherwise. This prevents $N$ from being negative. We then marginalise over the the expected number $N_{\rm{model}}$ of radio galaxies
 
\begin{equation}
P(data|OD)=\int P(N|OD,N_{\rm{model}}) \times P(N_{\rm{model}}) \rm{d}N_{\rm{model}} .
\end{equation}

\begin{figure}
\begin{center}
\setlength{\unitlength}{1mm}
\begin{picture}(150,65)
\put(-2,-10){\includegraphics{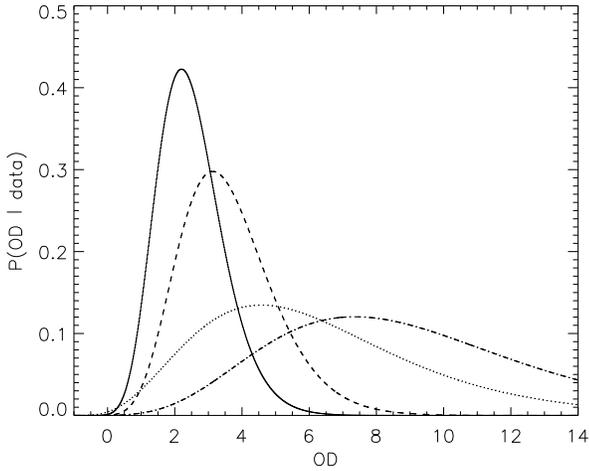}}
\end{picture}
\end{center}
{\caption[junk]{\label{fig:overdense} The probability of the radio galaxy overdensity given the data (incorporating both Poisson and model errors) for the TONS08 $z$=0.27 (solid line), the TONS08 $z$=0.35 (dashed line), the \citet{lac} $z$=0.28 (dot-dashed line) and the \citet{lac} $z$=0.35 (dotted line) redshift spikes.
}}
\end{figure}

We find an overdensity of $\delta_{\rm{gal}}\approx 2.27\pm^{0.51}_{0.21}$ for the $z$=0.27 redshift peak and $\delta_{\rm{gal}}\approx 3.41\pm^{0.72}_{0.49}$ for the z=0.35 redshift peak. The errors quoted are 68 per cent confidence limits. Fig.~\ref{fig:overdense} shows the probability distributions. 

\subsubsection{The Lacy sample}
\label{sec:lacy}

\begin{figure}
\begin{center}
\setlength{\unitlength}{1mm}
\begin{picture}(80,110)
\put(-5,120){\includegraphics{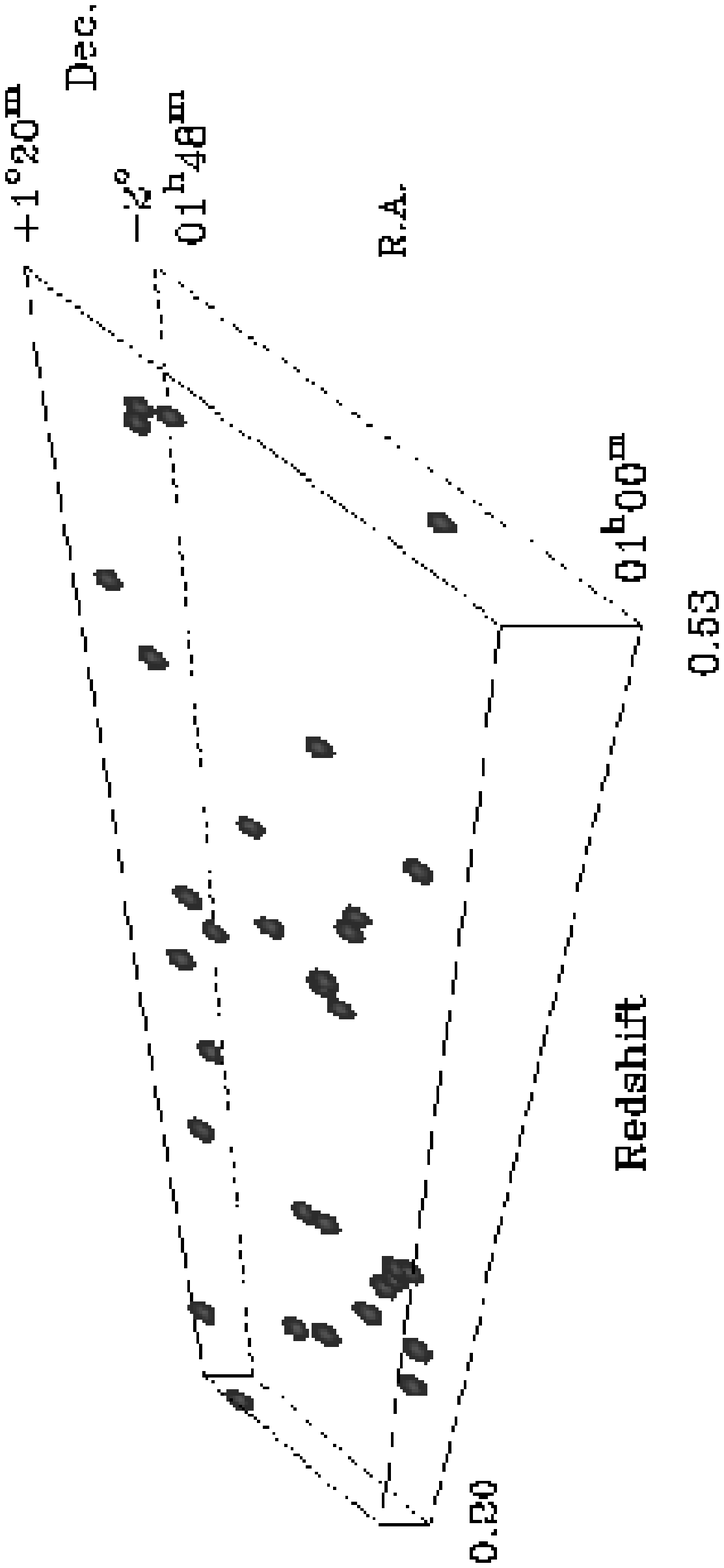}}
\put(0,70){\includegraphics{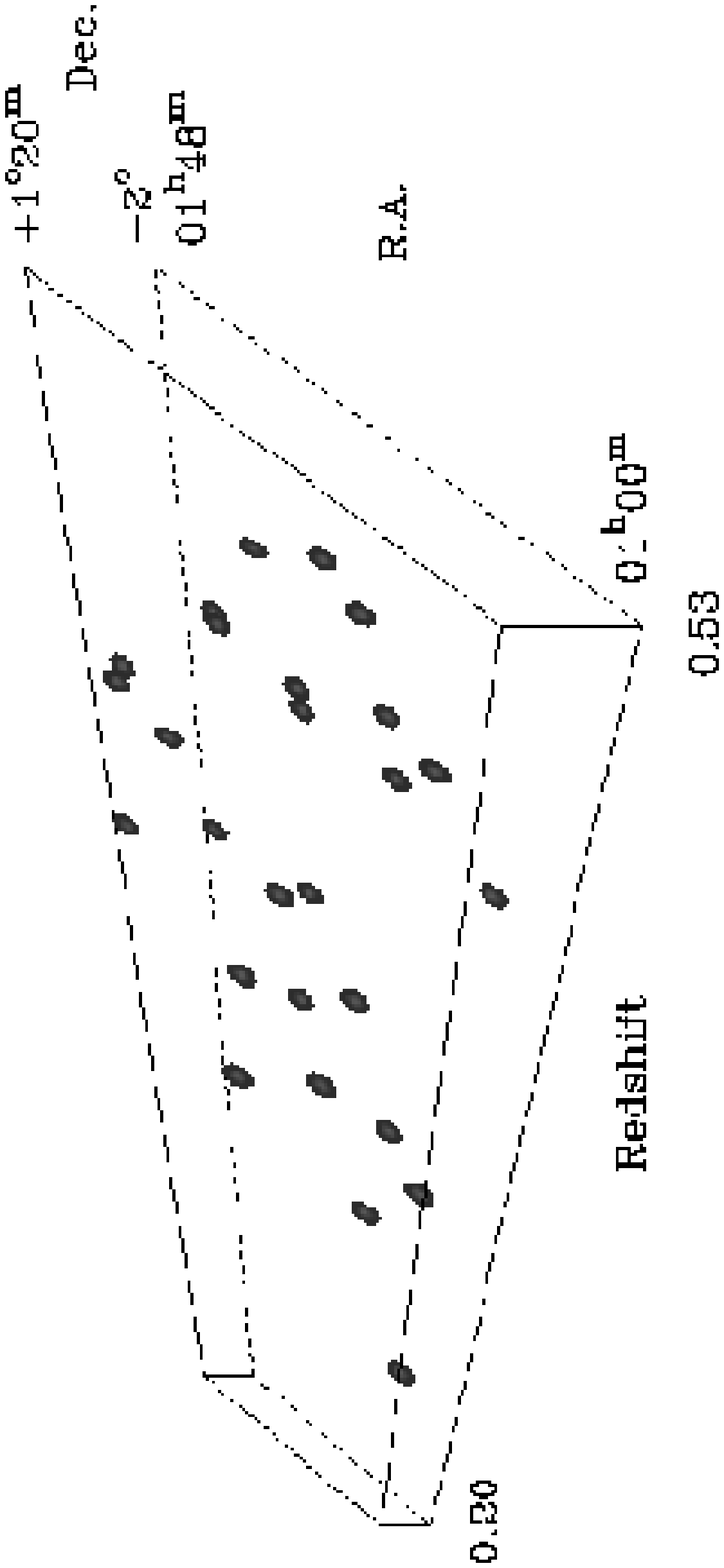}}
\put(0,40){\includegraphics{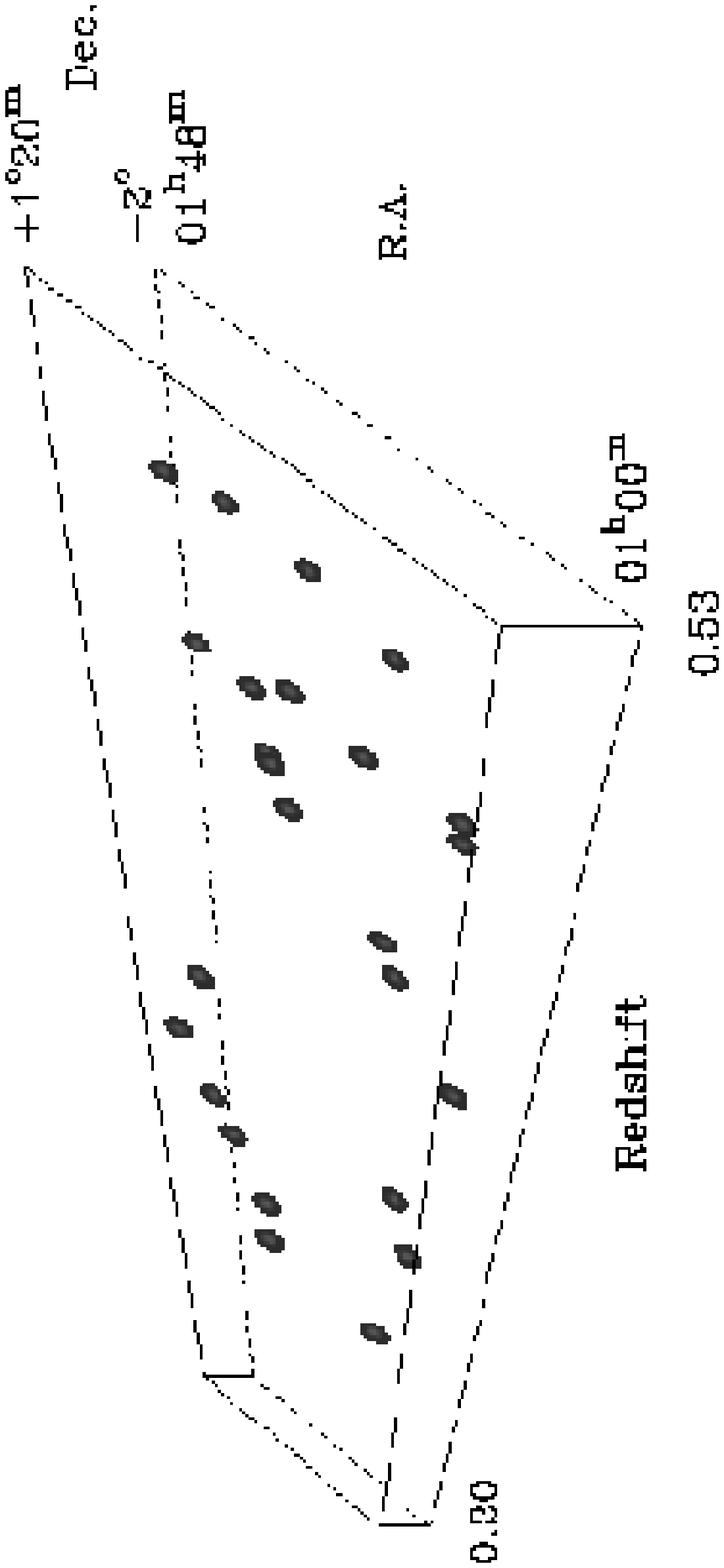}}
\end{picture}
{\caption[3-D distribution of TONS08 sample]{\label{fig:mark_3d} Representation of the three-dimensional distribution of the Lacy sample between $z$=0.2 and $z$=0.5 (top). Note the effect of the selection function in redshift as plotted in Fig.~\ref{fig:mark}. Also plotted below are two random samples with a selection function from the model redshift distribution (Fig.~\ref{fig:mark}). Similarly to TONS08, this sample looks significantly different to the random samples}}
\end{center}
\end{figure}

We also performed the above analysis for the \citet{lac} sample. This sample covers nearly twice the area of TONS08 in a different region of the sky and has a higher flux-density limit ($S_{\rm{1.4}}$=20 mJy) (see Table.~\ref{tab:samples}). We plot the 3-D distribution of this sample along with two random samples in Fig.~\ref{fig:mark_3d}. Clustering of radio galaxies can be clearly seen at $z\approx$0.3 and $z\approx$0.4, corresponding to the peaks in the redshift distribution (Fig.~\ref{fig:mark}). The random samples do not appear to show the same degree of clustering.
 
We find six radio galaxies within a 50 Mpc radius sphere at $z$=0.28 and four radio galaxies within a 50 Mpc radius sphere at $z$=0.36. In this case, we move the centre of the sphere in the RA direction to find the maximum number of radio galaxies within the sphere. This is because the Lacy survey is larger in extent in this direction ($\approx$ 300 Mpc at $z$=0.36) than the TONS08 survey. We are also confident in this case that we are tracing the entire super-structures. We predict there should be an average of 0.7 and 0.67 radio galaxies in the volume of the $z$=0.28 and $z$=0.36 spheres respectively (this is less than obtained for TONS08 due to the higher radio flux density limit of the Lacy sample).

We calculate the probability distribution of the overdensity $OD$ given the data for the two redshift spikes in the Lacy sample (Fig.~\ref{fig:overdense}). We find an overdensity of $\delta_{\rm{gal}}\approx 7.57\pm^{2.01}_{1.13}$ for the $z$=0.28 redshift peak and $\delta_{\rm{gal}}\approx 4.97\pm^{2.21}_{0.72}$ for the z=0.36 redshift peak. The errors quoted are 68 per cent confidence limits. We note here (Fig.~\ref{fig:overdense}) that the errors are much larger. There is a much higher probability that these spikes correspond to lower overdensities than they appear which is why, perhaps, not much attention has been drawn to them previously.

\subsection{Finding super-structures from radio galaxy optical surveys}

\begin{figure}
\begin{center}
\setlength{\unitlength}{1mm}
\begin{picture}(150,65)
\put(-2,-10){\includegraphics{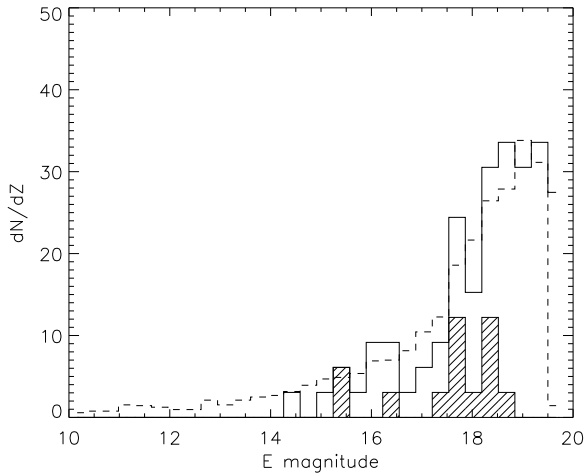}}
\end{picture}
\end{center}
{\caption[junk]{\label{fig:rmag} Histogram of the apparent magnitudes of the TONS08 sample (solid line) against a background level (dotted line) obtained from a similar sample in a larger $\sim$ 13-times (but then normalised) area of sky. The shaded area shows the 13 objects that are members of the $z$=0.27 super-structure.  
}}
\end{figure}

Fig.~\ref{fig:rmag} shows the distribution of apparent magnitudes in the TONS08 sample. A ``background'' level is also plotted from a sample selected in the same way over a much ($\sim$ 13-times) larger area of sky (but normalised down to the same sky area). There are no very significant spikes in the magnitude distribution. This is because objects at a given redshift have a large spread in absolute magnitude (Fig.~\ref{fig:rz}) and therefore any structures will become smeared out. This shows that redshift surveys are required to pick up significant super-structures.

\normalsize

\subsection{Angular clustering}
\label{sec:ang}

To test for any angular clustering we applied two-dimensional Kolmogorov-Smirnov (K-S) tests to our sample \citep{pre}. We compared the 2-D distribution of radio galaxies in significant super-structures (Sec.~\ref{sec:sig}) with the distribution of samples randomly distributed over the selection volume. To obtain a randomly distributed sample for the larger super-structure samples, we distributed the sample randomly over the sky area of TONS08. For the samples that are defined to be within a 50-Mpc-radius sphere, we did this by randomly distributing the sample within the volume of the 50 Mpc radius sphere and de-projected this back down to a 2-D angular distribution. 
The Kolmogorov-Smirnov $D$ is defined as the maximum value of the absolute difference between the two cumulative distributions. This is generalised to 2-D by considering the fraction of data points in each of four quadrants around a given point \citep{pre} and taking the maximum difference to define $D$. 

We compare the real distribution with a random distribution to find the value of $D$. To calculate the probability that the real distribution is picked from a different distribution to that of a random distribution, we also calculate $D$ for two random distributions. We do this 10000 times and calculate the fraction of random values of $D$ which are larger than that of the real $D$ to obtain the probability of the random $D$ being larger than the real $D$ (see Table.~\ref{tab:ang}). If this probability is small, the distribution is significantly different to a random distribution.

\begin{figure*}
\begin{center}
\setlength{\unitlength}{1mm}
\begin{picture}(80,70)
\put(-50,-10){\includegraphics{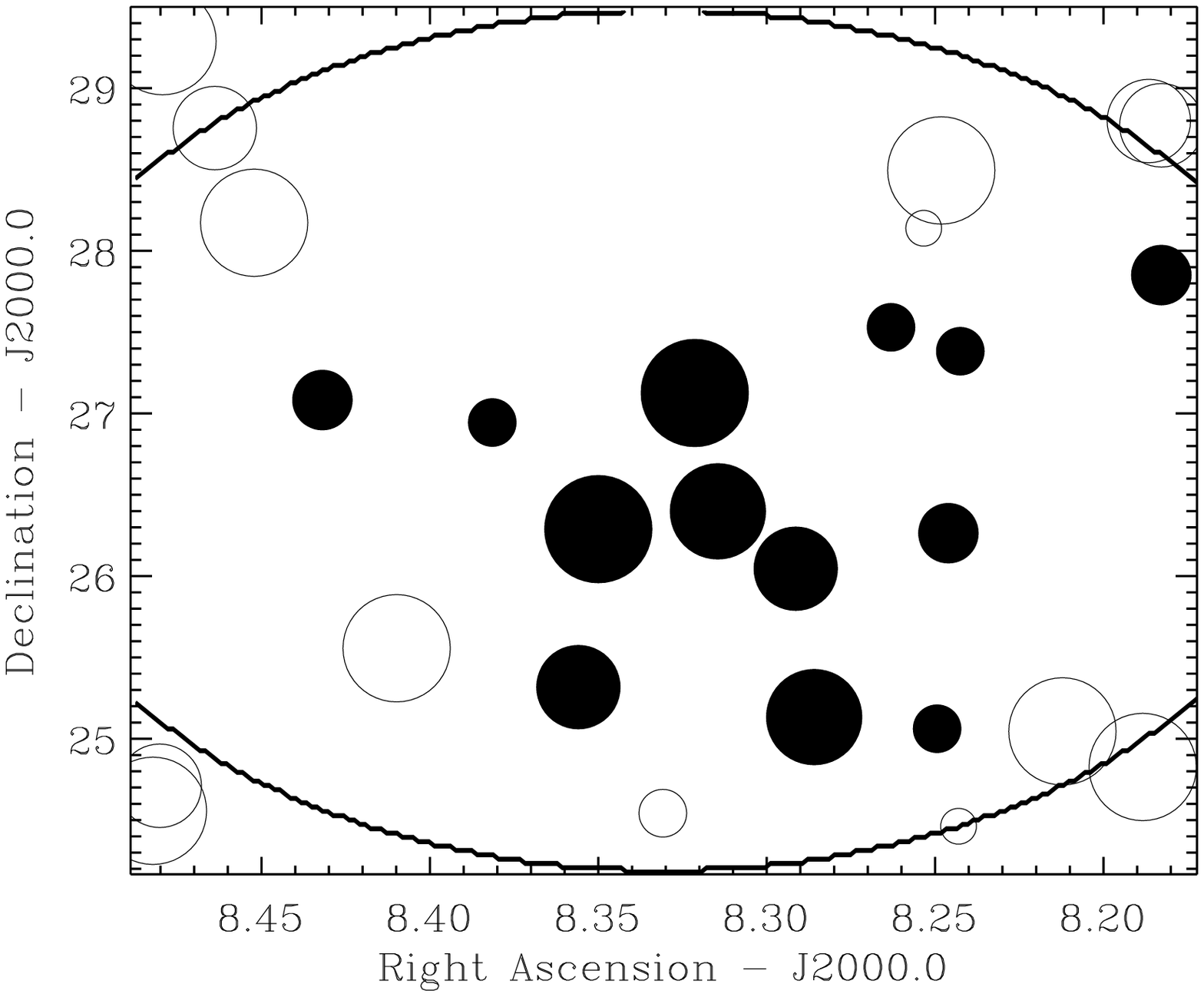}}
\put(30,-10){\includegraphics{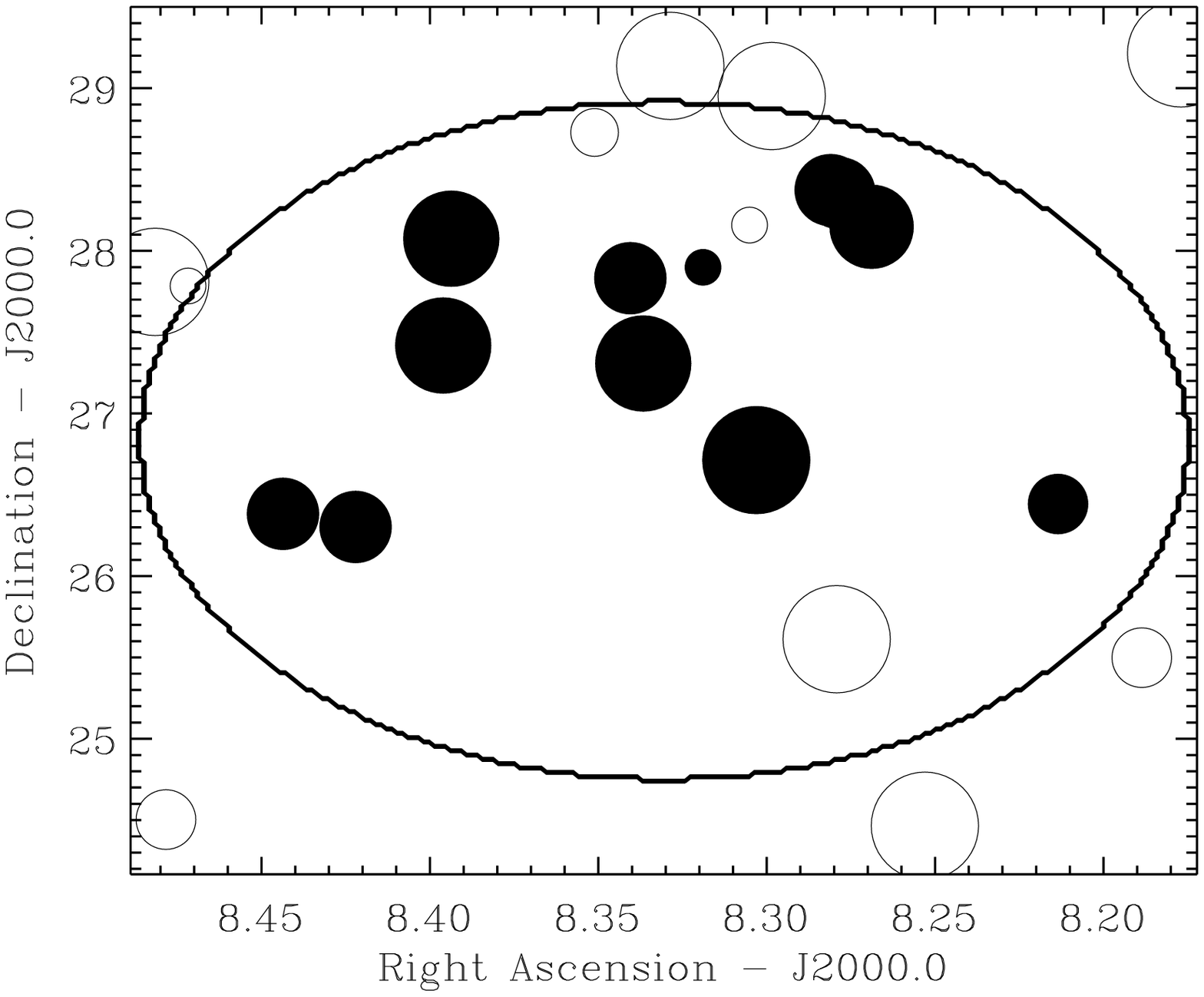}}
\put(-25,65){\textbf{$z$=0.27 Super-Structure}}
\put(55,65){\textbf{$z$=0.35 Super-Structure}}
\end{picture}
\end{center}
{\caption[2D distributions]{\label{fig:2d} The angular distribution of the $z$=0.27 (left) and the $z$=0.35 (right) super-structure members. The TONS08 survey is delimited by the axes surrounding each plot. Note, however, the members of the super-structures (filled circles) are restricted to a 50-Mpc-radius sphere whose angular projection is overplotted. We also plot the larger number of candidate super-structure members as described in Sec.~\ref{sec:sig}. Radio galaxies are scaled linearly according to their redshift (lower redshift objects are larger). The redshifts in the 50 Mpc spheres lie between $z$=0.26 and $z$=0.281 for the $z$=0.27 super-structure (left) and between $z$=0.344 and $z$=0.366 for the $z$=0.35 super-structure (right).
}}
\end{figure*}

We find very weak evidence for angular clustering in both the super-structures but not within the entire sample. The 2-D distributions are shown in Fig.~\ref{fig:2d}. Evidence for angular clustering, at least in the restricted set of super-structure members, include an apparent northern edge at DEC $\approx$28$^{\circ}$ in the $z$=0.27 super-structure and an apparent edge at DEC $\approx$26$^{\circ}$ in the $z$=0.35 super-structure. Edges to the distribution can also be perceived in Fig.~\ref{fig:3d}. The symbols representing radio galaxies are scaled with redshift to check there is no systematic velocity shift across the fields. No such effect is statistically significant but we note hints that the lower redshift objects in both superstructures are more tightly clustered than the whole sample. We have found very tentative evidence for angular clustering in the samples corresponding to the 50 Mpc spheres. This suggests that both these super-structures are real. 

\begin{table}
\begin{center}
\begin{tabular}{llll}
\hline\hline
Redshift & No. Galaxies & $D$ & Probability \\
\hline
0.27        & 13 & 0.38 & 0.43 \\
0.233-0.285 & 27 & 0.24 & 0.66\\
0.35        & 12 & 0.40 & 0.35 \\
0.330-0.367 & 23 & 0.31 & 0.37\\
0.0-0.5     & 84 & 0.16 & 0.58 \\
\hline\hline
\end{tabular}
{\caption[Table 2]{\label{tab:ang} Results of two-dimensional K-S tests for the spatial distribution of the significant super-structures (for both the larger sample and the sample which includes all radio galaxies within a 50 Mpc sphere) and the whole TONS08 sample. $D$ is defined as the maximum value of the absolute difference between the data and a random cumulative distribution. The Probability is that of a random-random value of $D$ being greater than the calculated value of $D$.
}}
 \end{center}
 \end{table}

\normalsize

\subsection{The geometry of the super-structures}
\label{sec:geom}

\begin{figure*}
\begin{center}
\setlength{\unitlength}{1mm}
\begin{picture}(150,120)
\put(-10,-15){\includegraphics{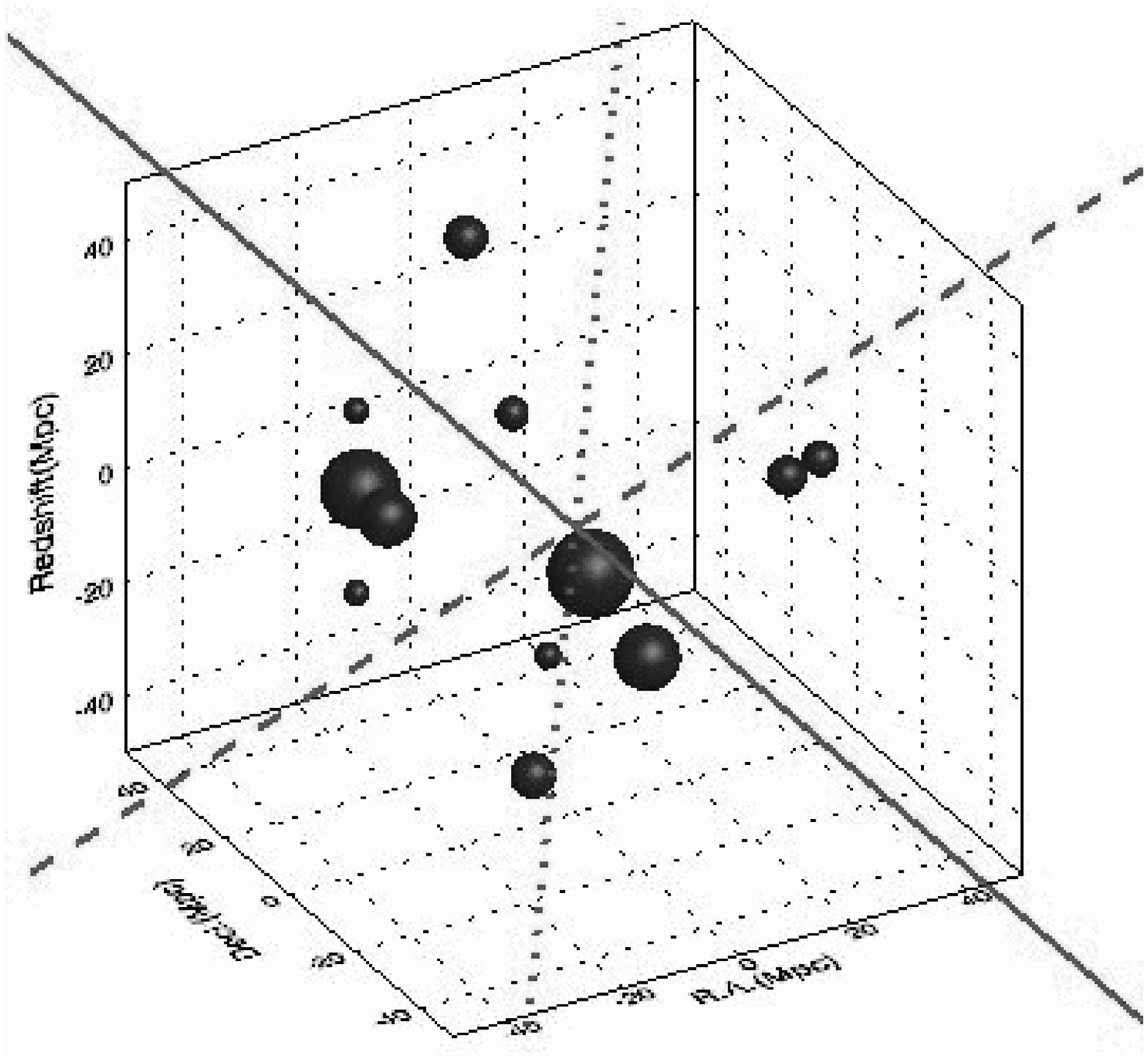}}
\put(50,-15){\includegraphics{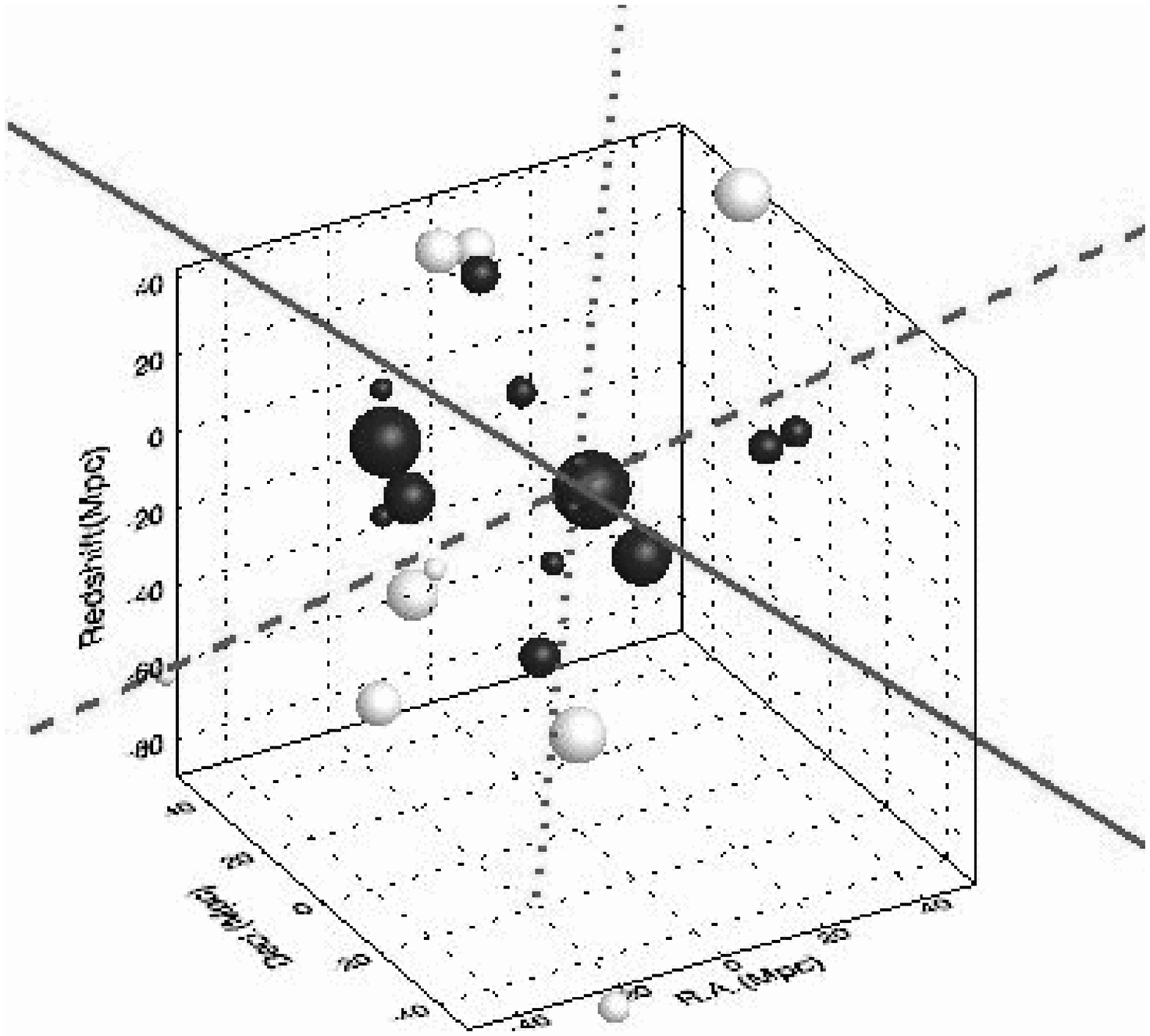}}
\put(100,-15){\includegraphics{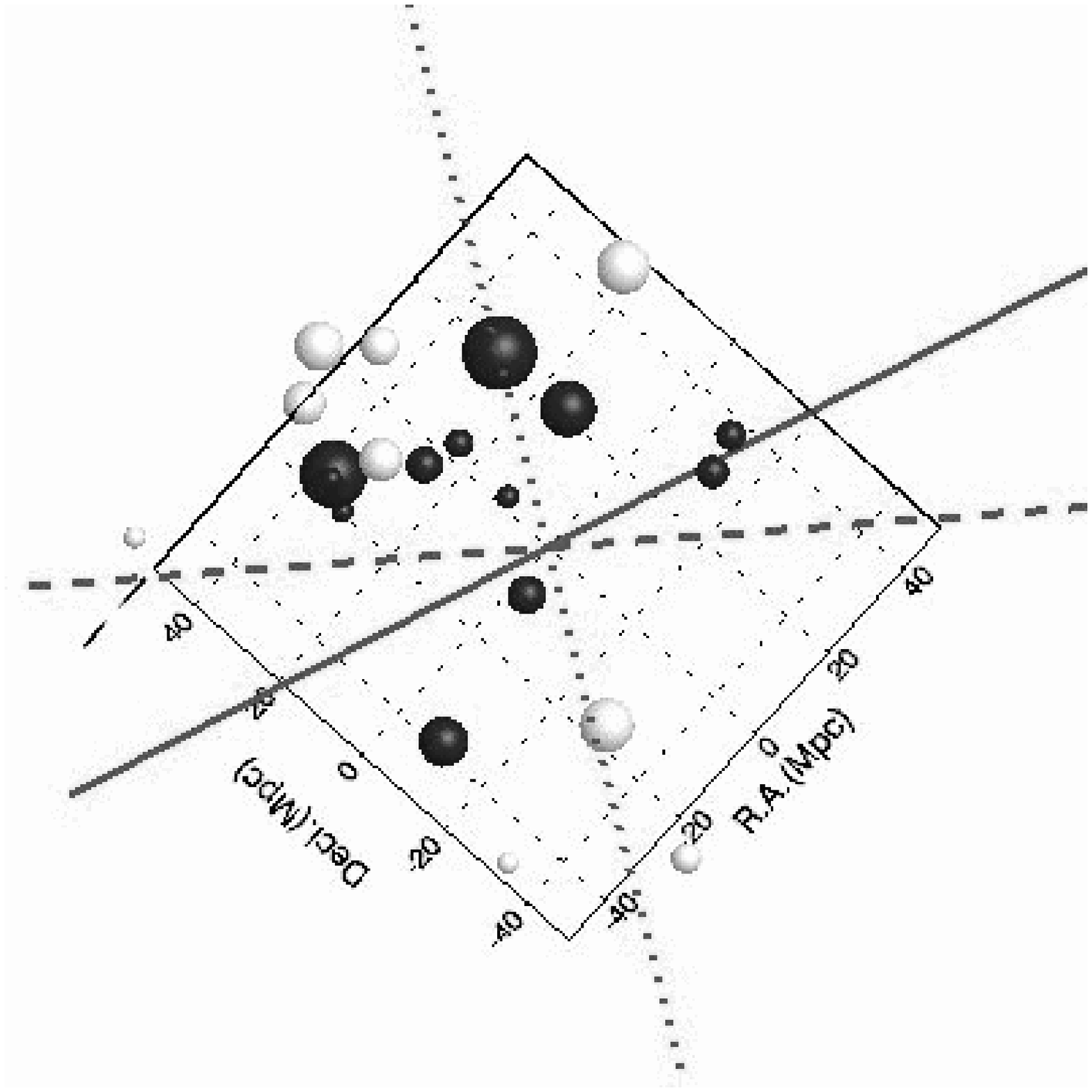}}
\put(-10,45){\includegraphics{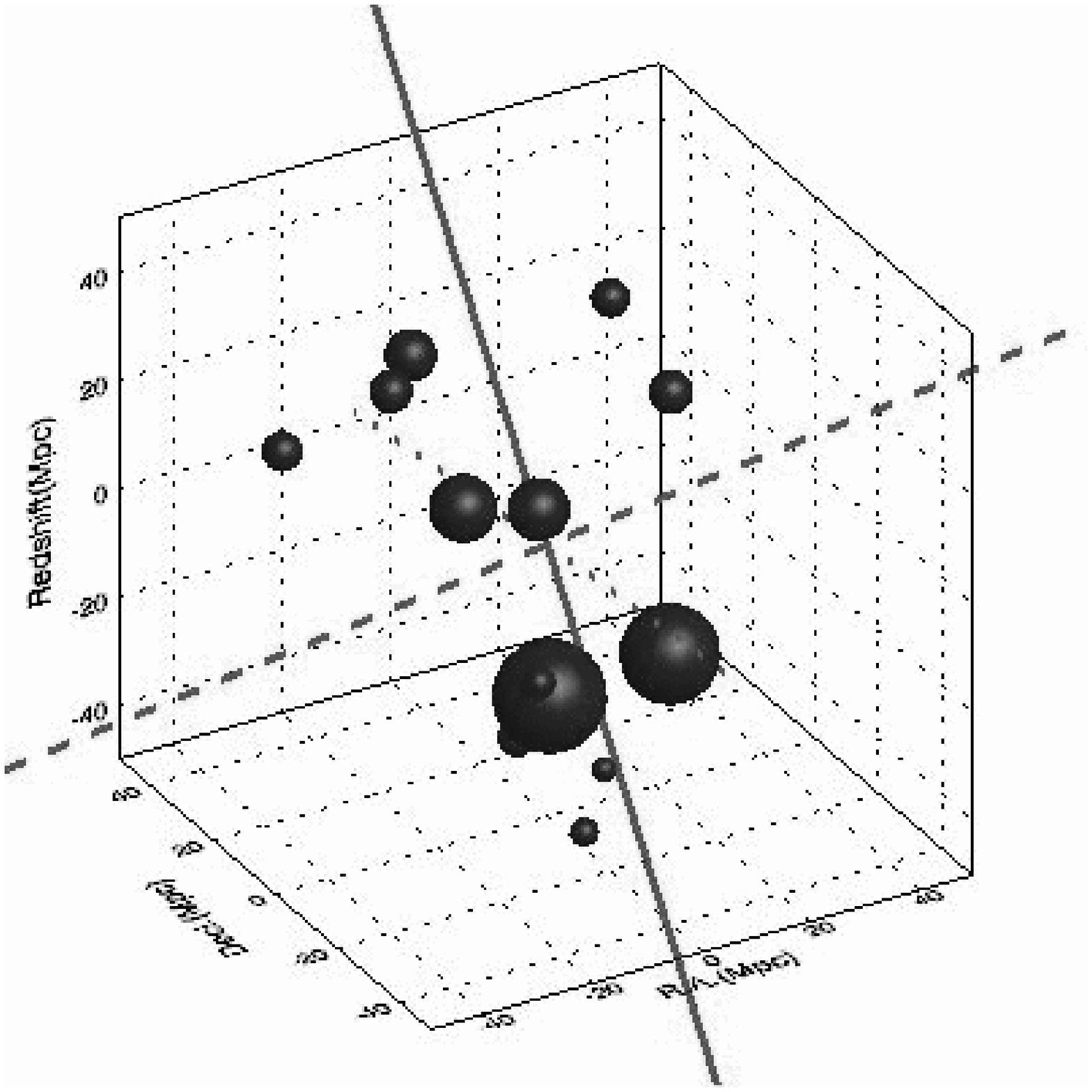}}
\put(50,45){\includegraphics{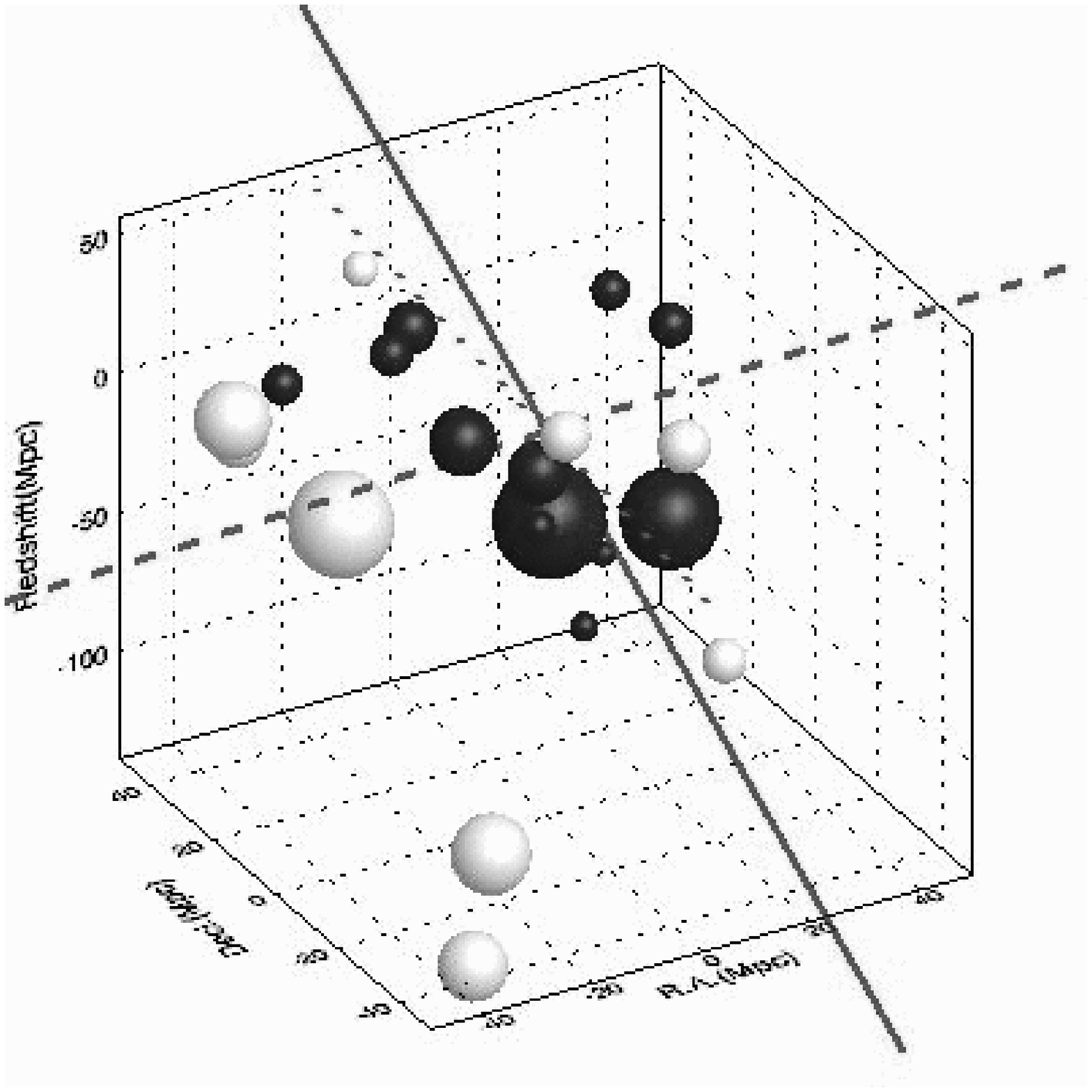}}
\put(100,45){\includegraphics{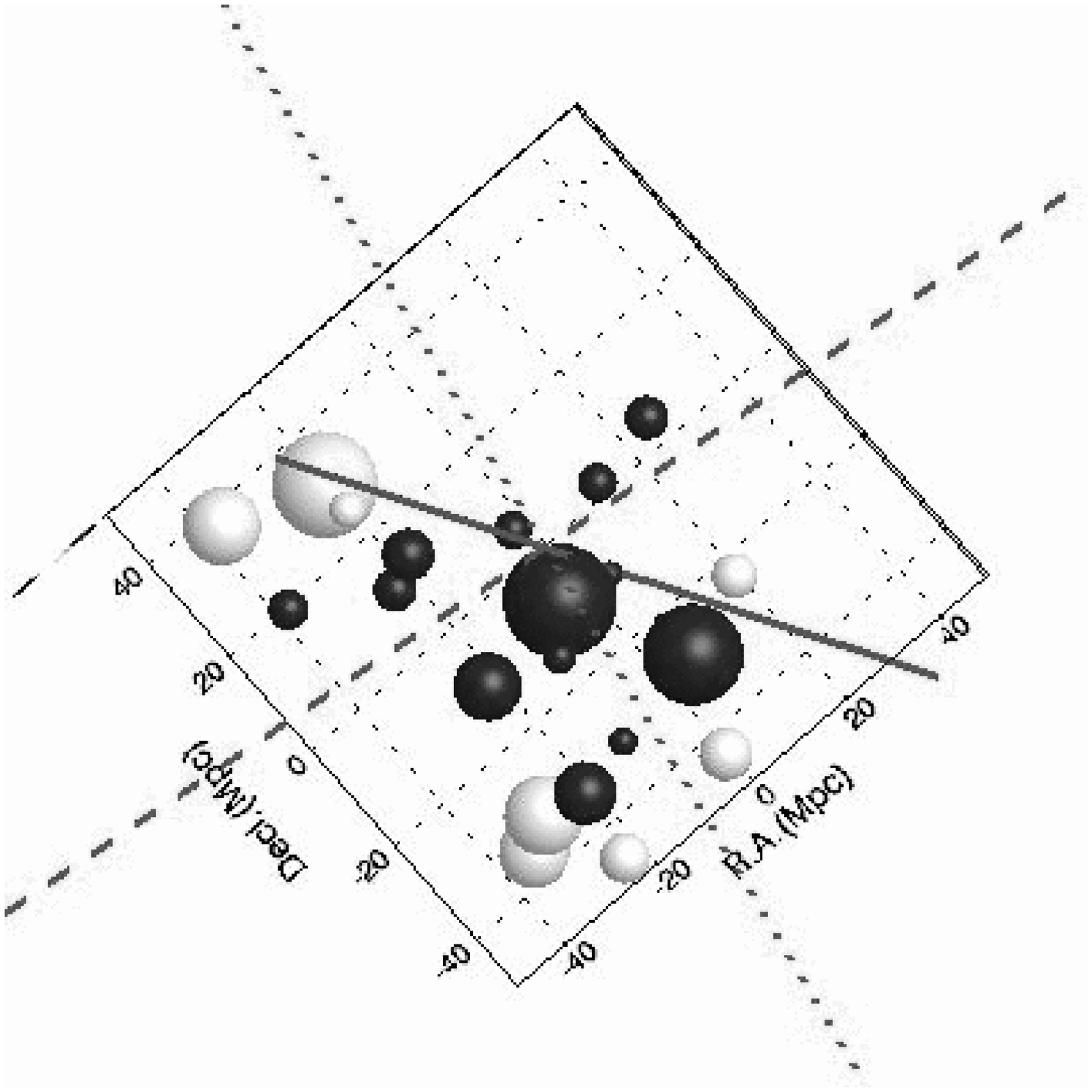}}
\put(50,115){\textbf{$z$=0.27 Super-Structure}}
\put(50,55){\textbf{$z$=0.35 Super-Structure}}
\end{picture}
\end{center}
{\caption[junk]{\label{fig:sc_pca} The spatial distribution of the $z$=0.27 super-structure (top) and the $z$=0.35 super-structure (bottom). The spheres representing the radio galaxies are scaled logarithmically with their radio flux densities. The solid line shows the first principal component. The dashed and dotted lines show the second and third principal components respectively (see Table.~\ref{tab:sc_pca} for coefficients). The figures on the left show the radio galaxies defined as being within 50 Mpc of the centre of the super-structure. The central figures additionally show the larger samples (white spheres). The figures on the right show both the larger (white spheres) and smaller samples (black spheres) with the principal axes de-projected in the $z$-direction to show how they would appear on the sky (i.e. Fig.~\ref{fig:2d}. The axes are plotted in co-moving distance (Mpc) so the boxes have sides of $\approx$100 Mpc (except the central figures which have a larger extent in the $z$ direction).
}}
\end{figure*}

The spatial extents of our $z$=0.27 and $z$=0.35 super-structures are at least $\approx$ 80 $\times$ 100 $\times$ 100 $\rm{Mpc}^3$ and 100 $\times$ 100 $\times$ 100 $\rm{Mpc}^3$ respectively ($\Delta$ RA $\times \Delta$ DEC $\times \Delta z$ in co-moving coordinates). These are limits due to selecting all radio galaxies within 50 Mpc of the centre of each super-structure. 
We used Principal Component Analysis (PCA) \citep{mh} to quantify whether the radio galaxies trace a spherical or more filamentary distribution.  PCA seeks the axis from which all the points have the smallest variance. The direction of this axis forms the first principal component (the first eigenvector). The next principal component is found by minimising the variance in a direction orthogonal to the first principal component. This process is continued to form three orthogonal axes. We convert the spatial and redshift coordinates to co-moving distances from the centre of the super-structure, making the parameters in correct proportion to each other.
The spatial distributions of radio galaxies in the two super-structures and their principal components are shown in Fig.~\ref{fig:sc_pca}. The eigenvalues are the variances of each principal component. We can therefore find the preferred direction of the spatial distribution (the first principal component) and quantify how much of the variance this accounts for (the eigenvalues). If the eigenvalues are all fairly equal in size this indicates a fairly spherical distribution. Conversely, if one eigenvalue is much larger than the others, this indicates a filamentary structure in the direction of the corresponding principal component. The results in Table.~\ref{tab:sc_pca} show a slight elongation for the $z$=0.27 super-structure. The direction of the eigenvectors shown in Fig.~\ref{fig:sc_pca} shows that this is mainly in the $z$-direction. 

Although we are performing the analysis on 50 Mpc radius spheres, there may be a bias introduced in the preferred direction of the PCA analysis. This is because the area of the survey corresponds to only about 80 Mpc in the RA direction for the $z$=0.27 super-structure. This effect will be even more pronounced for the larger samples. To take the survey volume into account and calculate the significance of any elongation, we performed a Monte-Carlo analysis. 

We calculated 100000 random realisations of the same number of objects within a 50 Mpc radius sphere (and within the redshift limits for the larger samples) for each super-structure. For each principal component, we found the percentage of these realisations whose eigenvalues were as far or greater from the mean random eigenvalue as the real eigenvalue. The results are shown in Table.~\ref{tab:sc_pca}. For the first principal component of the $z$=0.27 super-structure, only 2.5 per cent of the random realisations had eigenvalues as large as, or larger than, that measured. The geometry of the survey cannot, therefore, be responsible for this high eigenvalue.

We have not, however, corrected for redshift distortion effects.  If the super-structure is in-falling, this could make the super-structure appear less elongated in the $z$-direction. This could mean that the real elongation is even more pronounced. We neglect redshift space distortions here (as they will only make the results more extreme) but see Sec.~\ref{sec:bias} for a more thorough treatment.


\scriptsize
\begin{table}
\begin{center}
\begin{tabular}{lllll}
\hline\hline
Redshift    &       & PC1  & PC2   & PC3 \\
\hline
0.27        & Prop. &0.637 & 0.191 & 0.172 \\
            &random Prop &0.369 & 0.370 & 0.260 \\ 
            & MC    &2.5\% & 20.7\%& 39.7\% \\
0.233-0.285 & Prop. &0.669 & 0.196 & 0.135 \\
            &random Prop &0.717 & 0.182 & 0.101 \\ 
            & MC    &8.6\% & 40.5\%& 11.1\% \\
0.35        & Prop. &0.444 & 0.370 & 0.186 \\
            &random Prop &0.308 & 0.345 & 0.347 \\ 
            & MC    &80.8\% & 57.1\%& 20.9\% \\
0.330-0.367 &Prop.  &0.474 & 0.304 & 0.222 \\
            &random Prop &0.413 & 0.386 & 0.200 \\ 
            & MC    &19.3\% & 41.1\%& 21.0\% \\ 
\hline\hline
\end{tabular}
{\caption[junk]{\label{tab:sc_pca} The proportion of the total variance (Prop.) accounted for by the 1st (PC1),2nd (PC2) and 3rd (PC3) principal components for the $z$=0.27 and the $z$=0.35 super-structures as well as the larger samples. Also shown is the proportion of the total variance for random realisations of the sample (random Prop.), and the percentage of random realisations with the measured value or further from the mean (MC). The coefficients of the principal components are: PC1=[-0.17,0.29,0.94],PC2=[-0.97,0.12,-0.21] and PC3=[0.17,0.95,-0.26] for the $z$=0.27 super-structure and PC1=[0.64,-0.24,-0.73],PC2=[0.55,-0.53,0.65] and PC3=[0.54,0.82,0.20] for the $z$=0.35 super-structure
}}
 \end{center}
 \end{table}

\normalsize

\subsection{Implications of the radio galaxy overdensity}
\label{sec:bias}

We have established that the $z$=0.27 super-structure is significant and the $z$=0.35 super-structure probably is. We now attempt to calculate the probability of the TONS08 survey intercepting such structures given the cosmological parameters and different degrees of radio galaxy bias.

We define the bias factor $b$ as the ratio of the radio galaxy overdensity $\delta \rho_{\rm{g}} / \rho_{\rm{g}} $ to the underlying matter overdensity $\delta \rho_{\rm{m}} / \rho_{\rm{m}} $

\begin{equation}
\frac{\delta \rho_{\rm{g}}}{\rho_{\rm{g}}}=b\ \frac{\delta \rho_{\rm{m}}}{\rho_{\rm{m}}}.
\end{equation}

\noindent We assume the local value for the radio galaxy bias factor is $b$=1.65$\pm0.15$. This is determined by multiplying the relative bias factor of radio galaxies to optical galaxies (1.9/1.3=1.5$\pm$0.09 calculated by \citealt{pd}) with the bias factor of optical galaxies relative to the dark matter (1.1$\pm$0.08 calculated by \citealt{lah}) and adding the errors in quadrature.

\subsubsection{Press-Schechter approach}

If the bias factor were $\sim$1, the super-structures could be treated as non-linear collapsing objects and we could then use a Press-Schechter approach. Sec.~\ref{sec:geom} shows the radio galaxies in both super-structures trace an approximately spherical distribution which is an assumption in this approach. To apply the Press-Schechter approximation, we followed the method of \citet{ste}. Assuming some value for the bias factor $b$, we calculated the mass of each super-structure given by $M = <\rho>V(1+\delta_{\rm{m}})$ where $<\rho>$ is the mean density of the Universe, $V$ is the co-moving volume and $\delta_{\rm{m}}$ is the mass overdensity. We find a mass of $M\approx7\times10^{17}\rm{M_{\odot}}$ for each super-structure. We then calculated the probability of observing a peak this high in the survey volume in a $\Lambda$CDM Universe using the Press-Schechter approximation given in \citet{ste}. For a bias factor, $b$=1, we obtained a probability of $<10^{-11}$. We conclude that, like other radio galaxies, the TONS08 radio galaxies are not unbiased tracers of mass.

\subsubsection{Quasi-linear structure formation theory}

As we increase the bias factor, the radio galaxy overdensity corresponds to a smaller mass overdensity. Structures go non-linear at overdensities of approximately 1.7. Therefore, as the bias increases, the super-structures are still in the quasi-linear regime (albeit only just) and the Press-Schechter method is not applicable. As discussed above, we have good reason to expect the bias factor to be greater than 1 as locally, it has been measured to be $\approx$1.65 \citep{pd}. We therefore adopt the following method.

We model the height of the cosmic power spectrum at the relevant scale. We consider the dimensionless power spectrum, $\Delta^{2}(k)$ as this gives us the dimensionless spectrum of density fluctuations per ln $k$. As we don't know the full extent of the structures, we calculate this for the scale 50 Mpc (radius) ($k$=0.126). The normalisation of the power spectrum is fairly well known down to wave numbers of approximately 0.028 \citep{pd}. 

We calculate the power spectrum at redshift $z$ and for a wavenumber $k$ using

\begin{equation} 
\Delta^{2}(k)=\left(\frac{aG_{z}}{G_{0}}\right)^2 \left(\frac{t_{k}}{t_{8}}\right)^2 {\sigma_{8h^{-1}}}^2 \left(\frac{q}{q_{8}}\right)^{3+n},
\end{equation}

\noindent where $a=1/(1+z)$, $G_z$ is the growth factor at redshift $z$ \citep{eis}, $t_{\rm{k}}$ is the transfer function as given in \citet{bon}, $n$ is the index of the power spectrum and $q = k/\Gamma$. The effective wavenumber $k_{\rm{eff}}$ is defined as the wavenumber at which $\sigma_8$ measures power,

\begin{equation} 
k_{\rm{eff}}/h=0.172+0.011\left[\rm{ln}\left(\frac{\Gamma}{0.34}\right)\right]^2,
\end{equation} 

\noindent\citep{pea}. This then defines $q_{8}= k_{\rm{eff}}/\Gamma$ and a transfer function $t_{8}$ calculated at $k_{\rm{eff}}$. We then estimate the first three moments of the variance

\begin{equation} 
\sigma_{\rm{m}}^2=\int\Delta^2(k){\mid{f_k}\mid}^2 k^{2m-1}\rm{d}k, 
\end{equation} 

\noindent where $\mid{f_k}\mid$ is the Fourier transform of a 3-D top-hat filter function in which we set the radius to 50 Mpc and $m$ is the index of the moment. For the first moment, we obtain a value of $\sigma_{0}$ = 0.243. This is the rms density variation in 50 Mpc spheres. We can check this value is correct by the following rough calculation. Adopting the canonical form of the correlation function for galaxies, $\xi=(r/r_0)^{-1.8}$, and assuming $\sigma_{0}^2 \approx \xi$, we can calculate $\sigma_{50}^2=\sigma_{8h^{-1}}^2 \times (50 h/8)^{-1.8}$, where $\sigma_{8h^{-1}}$ = 0.94. We obtain the value $\sigma_{50}$ = 0.249 which agrees well with the above calculation.

The spectral parameters \citep{bar} are then calculated

\begin{equation} 
\nu=\frac{\delta_{\rm{m}}}{\sigma_0}, \gamma=\frac{\sigma_1^2}{\sigma_0\sigma_2}, R_\star=\sqrt{3}\frac{\sigma_1}{\sigma_2},
\end{equation} 

\noindent$\delta_m$ is estimated as before using $\delta_{\rm{m}}$=$\delta_{\rm{gal}}/b$. The rarer the density peak, the more it will have significant departures from a simple Gaussian random field due to non-linear growth by self-gravitation. We use a log-normal (LN) random field \citep{cj} to make some attempt at correcting the mass overdensity, $\delta_{\rm{m}}$ to a non-linear corrected mass overdensity $\delta_{\rm{nl}}$

\begin{equation} 
\delta_{\rm{nl}}=\log(1+\delta_{\rm{m}})+\frac{\sigma_{\rm{nl}}^2}{2},
\end{equation} 

\noindent where 

\begin{equation} 
\sigma_{\rm{nl}}^2=\log(\sigma_{0}^2+1) .
\end{equation} 

\begin{figure*}
\begin{center}
\setlength{\unitlength}{1mm}
\begin{picture}(150,70)
\put(20,55){{\textbf{$z$=0.27 Super-Structure}}}
\put(0,-5){\includegraphics{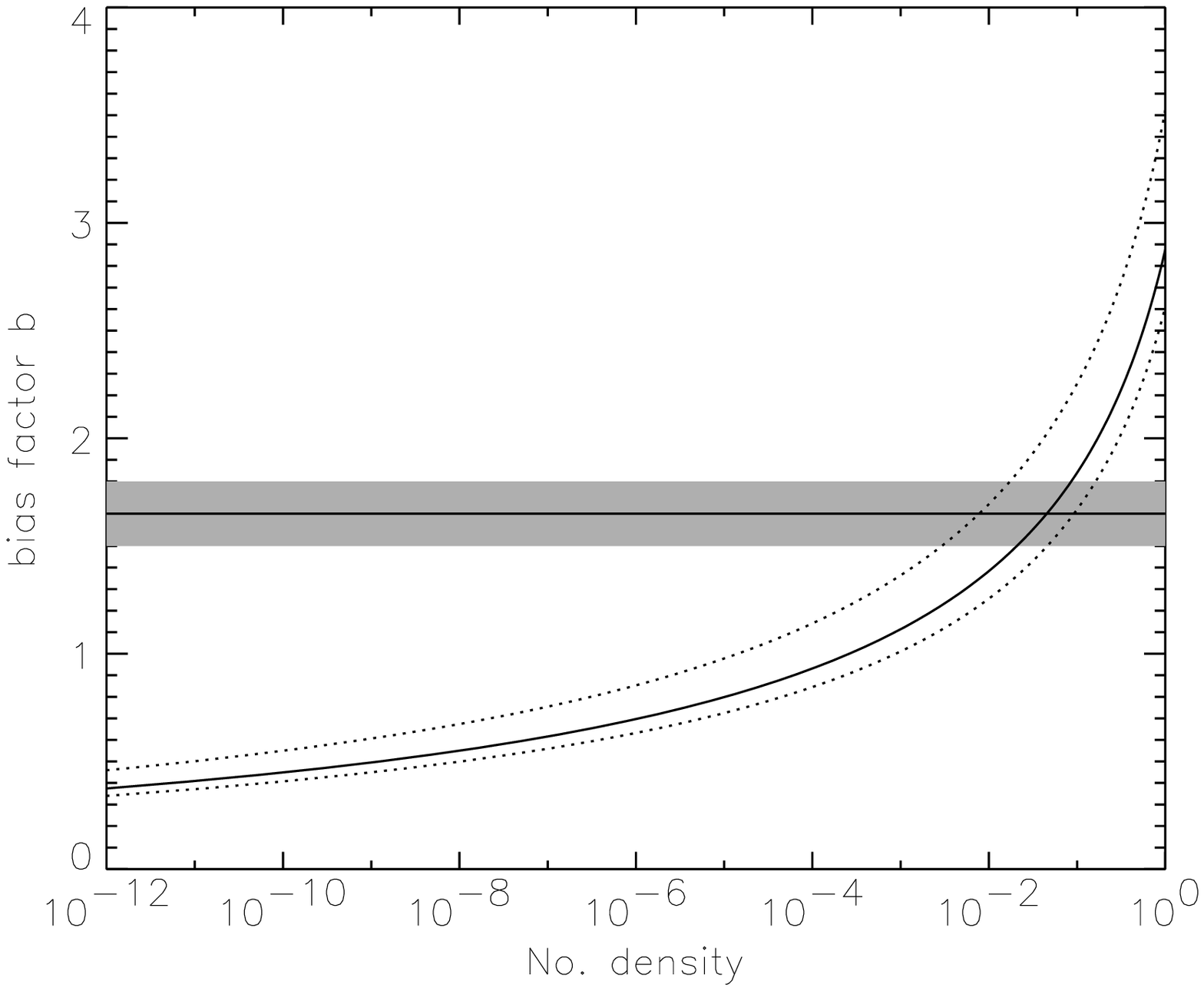}}
\put(90,55){{\textbf{$z$=0.35 Super-Structure}}}
\put(70,-5){\includegraphics{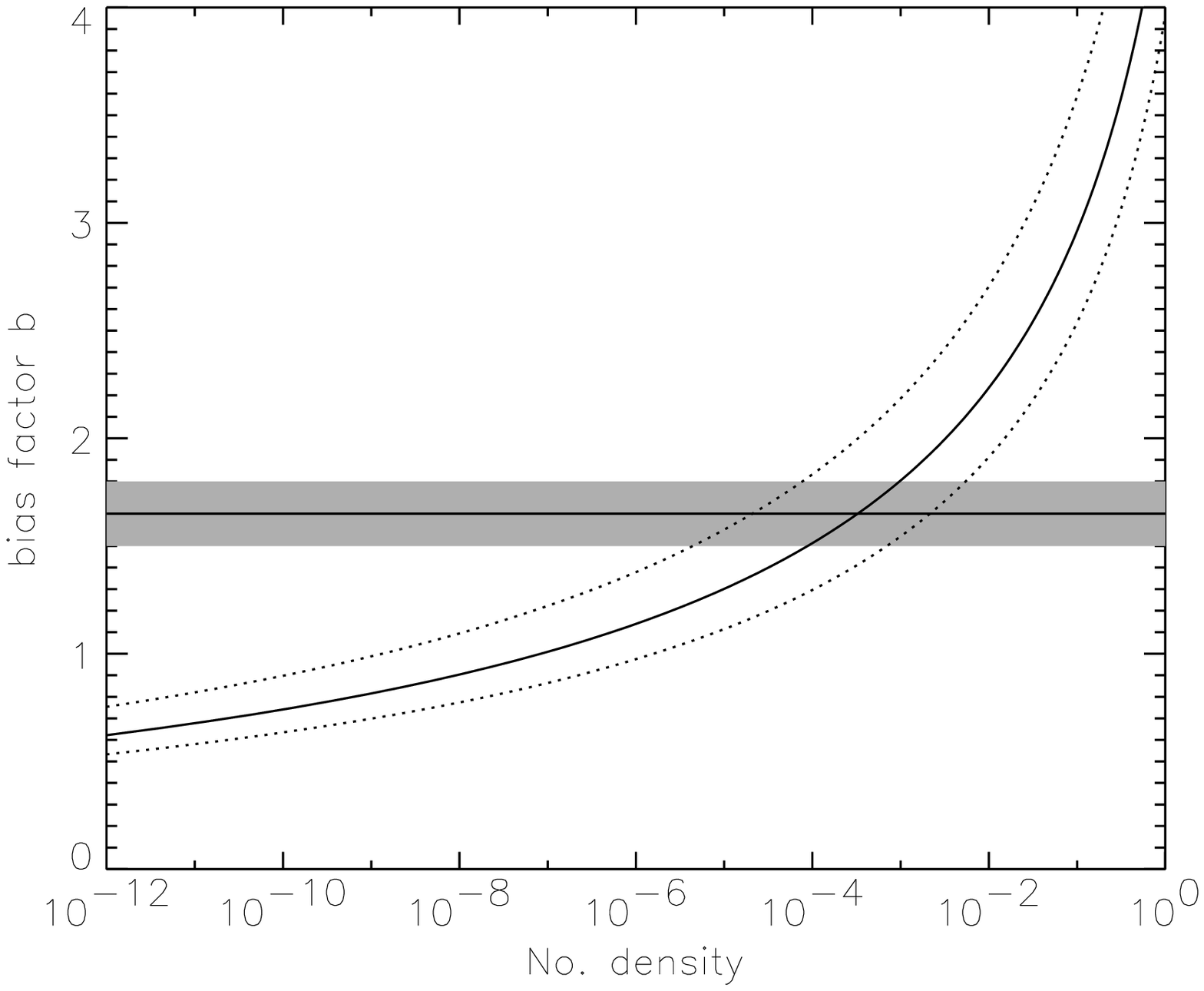}}

\end{picture}
\end{center}
{\caption[junk]{\label{fig:prob_bias} The probability of intercepting a super-structure of radius 50 Mpc and height associated with the galaxy overdensity of the $z$=0.27 super-structure (left) and $z$=0.35 super-structure (right) against the radio galaxy bias factor. $\sigma_{8h^{-1}}$ is assumed to be 0.94. Overplotted are the 68 per cent confidence limits of the galaxy overdensities which gives us an indication of the effects of Poisson sampling fluctuations. A local value for the radio galaxy bias parameter b=1.65$\pm$0.15 is plotted by horizontal lines with the regions within the error shaded. 
}}
\end{figure*}

The probability of intercepting a structure of height $\nu$ or greater was calculated by multiplying the co-moving volume of TONS08 ($1.5\times10^7 \rm{Mpc}^3$) by the cumulative number density of peaks higher than $\nu$. We used the fitting formulae of \citet{bar} (equations 4.3,4.4,4.5 and 4.11a). The results are shown in Fig.~\ref{fig:prob_bias}. As the results are very dependent on the normalisation of the power spectrum, we performed the analysis for $\sigma_{8h^{-1}}$=0.75 from combined 2dF/CMB measurements \citep{sel} and $\sigma_{8h^{-1}}$=0.94 from weak gravitational lensing measurements \citep{ref}. The results are shown in Fig.~\ref{fig:prob_bias} and Table.~\ref{tab:prob_bias}. 

\scriptsize
\begin{table*}
\begin{center}
\begin{tabular}{c|llll|llll}
\hline\hline
\multicolumn{9}{|c|}{$z$=0.27 Super-structure} \\
\hline\hline
&  \multicolumn{4}{|c|}{$\sigma_{8h^{-1}}$=0.73} & \multicolumn{4}{|c|}{$\sigma_{8h^{-1}}$=0.94}  \\
Model & \multicolumn{4}{|c|}{----------------------------} & \multicolumn{4}{|c|}{----------------------------}\\
Parameters & $b$=1.5 & $b$=1.65 & $b$=2 & $b$=4 & $b$=1.5 & $b$=1.65 & $b$=2 &$b$=4\\
\hline
$\delta_{\rm{m}}$           &1.51   &1.38  &1.14  &0.57 &1.51  &1.38  &1.14  &0.57\\
$\delta_{\rm{nl}}$          &1.02   &0.96  &0.86  &0.55 &1.05  &0.99  &0.88  &0.57\\
mass ($10^{17}\rm{M_\odot}$)&4.5    &4.4   &4.1   &3.4  &4.6   &4.4   &4.2   &3.5 \\
$\rm{\nu}$                  &5.18   &4.88  &4.35  &2.78 &4.13  &3.90  &3.48  &2.26\\
no. density	            &2.5E-4 &9.4E-4&9.1E-3&1.02 &0.02  &0.05  &0.17  &2.65\\
\hline\hline
\multicolumn{9}{|c|}{$z$=0.35 Super-structure} \\
\hline\hline
&  \multicolumn{4}{|c|}{$\sigma_{8h^{-1}}$=0.73} & \multicolumn{4}{|c|}{$\sigma_{8h^{-1}}$=0.94}  \\
Model & \multicolumn{4}{|c|}{----------------------------} & \multicolumn{4}{|c|}{----------------------------}\\
Parameters & $b$=1.5 & $b$=1.65 & $b$=2 & $b$=4 & $b$=1.5 & $b$=1.65&$b$=2 & $b$=4 \\
\hline
$\delta_{\rm{m}}$           &2.27   &2.07  &1.71  &0.85 &2.27  &2.07  &1.71  &0.85\\
$\delta_{\rm{nl}}$          &1.28   &1.21  &1.09  &0.71 &1.31  &1.24  &1.11  &0.74\\
mass ($10^{17}\rm{M_\odot}$)&5.1    &4.9   &4.6   &3.8  &5.1   &5.0   &4.7   &3.9 \\
$\nu$                       &6.79   &6.44  &5.77  &3.77 &5.38  &5.11  &4.59  &3.03\\
no. density	            &2.6E-8 &2.4E-7&1.1E-5&0.07 &8.8E-5&3.3E-4&3.2E-3&0.55 \\
\hline\hline
\end{tabular}
{\caption[junk]{\label{tab:prob_bias} Table showing the mass overdensity $\delta_{\rm{m}}$, the non-linear corrected mass overdensity $\delta_{\rm{nl}}$, the mass associated with this overdensity, the spectral parameter $\nu$ (the height of the field in units of the rms) and the expected number of peaks of height $\nu$ or greater in the survey volume for different values of radio galaxy bias and $\sigma_{8h^{-1}}$ for the two super-structures. 
}}
 \end{center}
 \end{table*}

\normalsize

Without taking into account non-linear effects, if we assume a relatively high value for $\sigma_{8h^{-1}}$ (0.94) and that the radio galaxy bias parameter is the same as for smaller scales in the local Universe (1.65), we have found an extremely rare (8$\sigma$) fluctuation (the $z$=0.27 super-structure). If we include a non-linear correction, the effective mass overdensity is reduced and corresponds to a much less extreme rare fluctuation of 3.9$\sigma$. After non-linear corrections, we calculate that we would only expect 0.05 such super-structures in our sample volume. These corrections are therefore very important. \citet{ste} use a different method for making non-linear corrections \citep{ber} which give similar results.

\subsubsection{Redshift space distortions}
\label{sec:zdistort}

In calculating the radio galaxy overdensity, we may need to account for redshift space distortions. As $\delta\rho/\rho$ is close to the critical value, we have until now assumed that the super-structure is just breaking away from the Hubble expansion. However, if it is more evolved and is collapsing, then the super-structure will appear more dense than it actually is. 

If we assume a maximum peculiar velocity $v_{\rm{pec}}$ towards the centre of the superstructure, we can work out the actual redshift of each radio galaxy and hence how many radio galaxies are actually outside the 50 Mpc radius sphere and calculate the new radio galaxy overdensity. 

The maximum radial infall velocity $v_{\rm{pec}}$ around a spherical perturbation of radius $R$ and interior average mass overdensity $\delta_{\rm{m}}$ is given by

\begin{equation}
v_{\rm{pec}}=\frac{1}{3} H_0 R f \delta_{\rm{m}} ,
\end{equation}

\noindent where f$\approx \Omega_{\rm{m}}^{0.6}$ \citep{lah91}. 

\begin{figure*}
\begin{center}
\setlength{\unitlength}{1mm}
\begin{picture}(150,70)
\put(20,55){{\textbf{$z$=0.27 Super-Structure}}}
\put(0,-5){\includegraphics{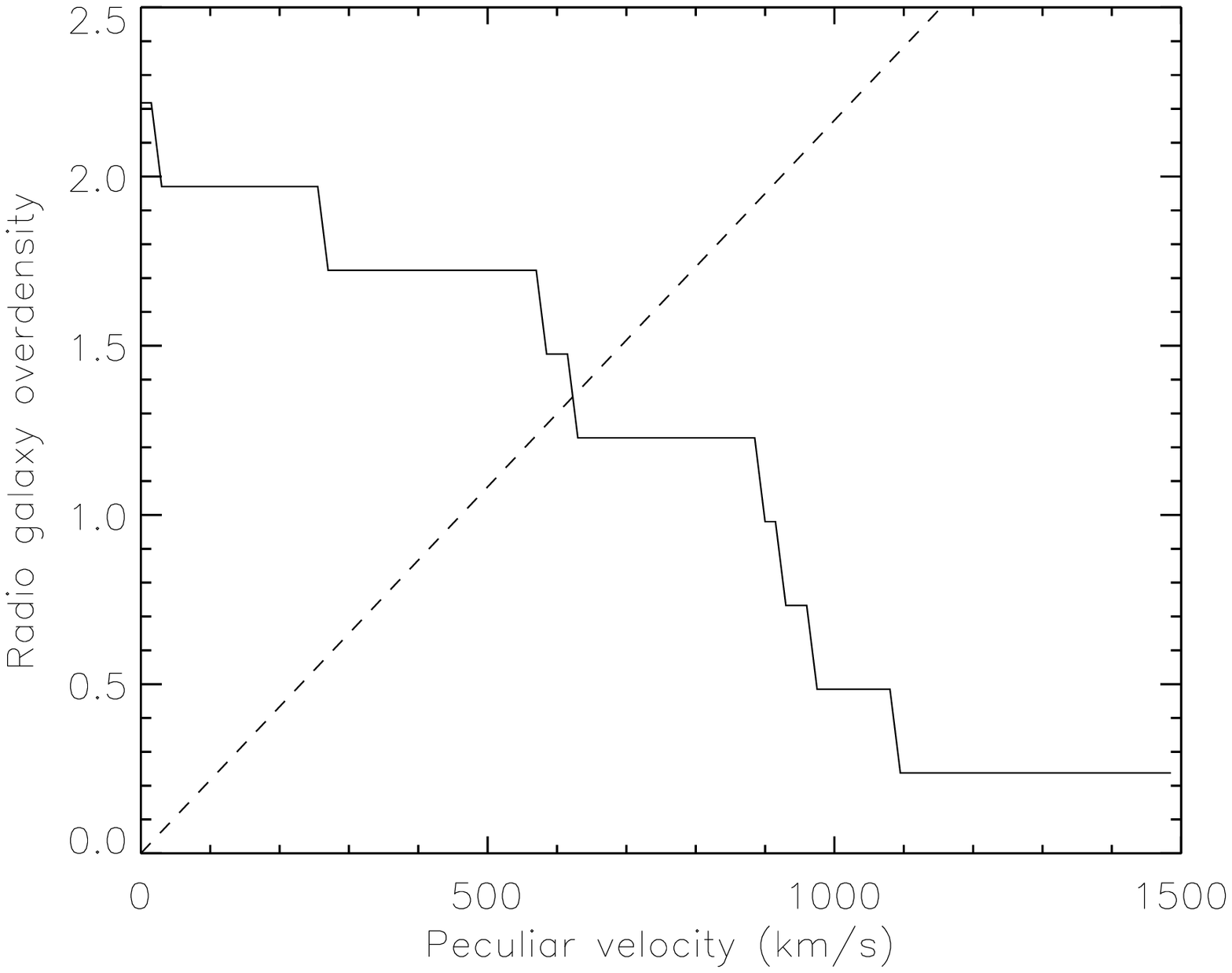}}
\put(90,55){{\textbf{$z$=0.35 Super-Structure}}}
\put(70,-5){\includegraphics{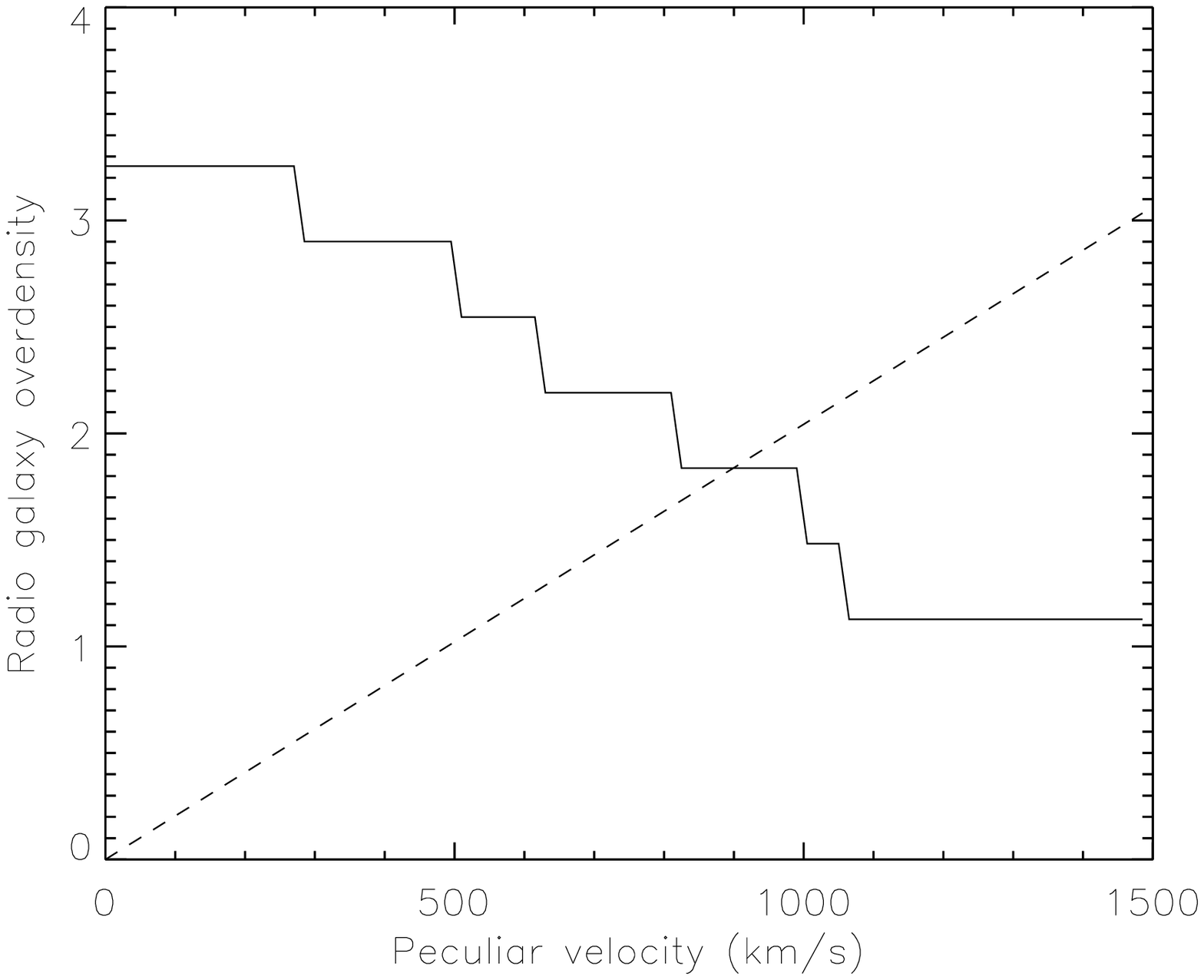}}

\end{picture}
\end{center}
{\caption[junk]{\label{fig:od_vpec} The redshift-space distortion corrected radio galaxy overdensity for different values of maximum peculiar velocity for both the $z$=0.27 (left) and $z$=0.35 super-structures. The results are shown for the data (solid line) and theory (dotted line). We chose a compromise radio galaxy overdensity by determining where these curves cross. 
}}
\end{figure*}

The problem in calculating this maximum infall velocity is that as we increase the value of the infall velocity, fewer radio galaxies fall within the 50 Mpc radius sphere. This in turn decreases the mass overdensity which will decrease the peculiar velocity (due to Equation 17). We therefore choose to find a compromise value for the redshift-space distortion corrected radio galaxy overdensity $\delta_{\rm{gal}}^{\rm{cor}}$ by calculating the radio galaxy overdensity given different $v_{\rm{pec}}$ for both the data and theory (from equation 17 and assuming a radio galaxy bias factor of $b$=1.65). We find a ``corrected'' radio galaxy overdensity of $\delta_{\rm{gal}}^{\rm{cor}}$=1.35 (10 radio galaxies) and 1.84 (8 radio galaxies) for the $z$=0.27 and 0.35 super-structures respectively (see Fig.~\ref{fig:od_vpec}). Of course, real collapsing systems will be more complex and hence have more complex velocity fields, so this is only a rough calculation.

Using these corrected radio galaxy overdensities (and assuming a radio galaxy bias factor $b$=1.65), the spectral parameter $\nu$=2.84 and 3.58, the non-linear corrected mass overdensity $\delta_{\rm{nl}}$=0.72 and 0.87, and the expected number of peaks of height $\nu$ or greater in the survey volume =0.87 and 0.12 for the $z$=0.27 and 0.35 super-structures respectively. Therefore, if these super-structures are collapsing, this has a huge effect on the results and it is not that surprising that we detect them in the TONS08 survey volume.

\section{DISCUSSION}
\label{sec:discussion}

\subsection{The $z$=0.27 super-structure}
\label{sec:discussion1}

\subsubsection{The geometry of the $z$=0.27 super-structure}

The spatial extent of the $z$=0.27 super-structure in TONS08 is $\approx$ 80 $\times$ 100 $\times$ 100 $\rm{Mpc}^3$ (see Sec.~\ref{sec:geom}).  A turn over in the matter power spectrum at these scales is still disputed \citep{mb}, mostly due to problems with window functions and sample incompletenesses. Recent results from the full 2dfQSO survey suggests no turnover in the power spectrum below scales of 210 Mpc \citep{out}. However, it is clear that our radio-selected super-structures are comparable to the largest structures found to date. For example \citet{bms} use optical clusters to find four super-clusters with a maximum extent of $\sim$80 Mpc. These structures, as well as most structures found to date, do not extend as far in all directions as TONS08 (although of course radio galaxies trace the TONS08 super-structures so sparsely, it is difficult for us to make firm statements). 

Sec.~\ref{sec:geom} makes an attempt to quantify the 3-D distribution of radio galaxies within the super-structures. The $z$=0.27 super-structure appears more elongated roughly in the z-direction. If we assume that the super-structure is in-falling, then redshift space distortions (see Sec.~\ref{sec:zdistort}) will make this an even more pronounced effect. \citet{zel} argued that pancakes are the first structures formed by gravitational clustering and are the dominant features arising from the first stages of non-linear gravitational clustering. They flatten out into sheets and then break up into filaments or form directly into sheets. The $z$=0.27 super-structure could therefore plausibly be elongated in a particular direction because of the way it is collapsing.
Alternatively, it is possible that the structure is bigger in all three dimensions, but the angular extent of the TONS08 survey means that we don't see all of it. If true, then this structure would become truly remarkable as it would correspond to a very high-sigma fluctuation\footnote{We have therefore conducted a new, wider angle, higher radio flux density survey (TOOT08w: Table.~\ref{tab:samples}) to further constrain the dimensions of the super-structure (Brand et al. in prep.).}.
It is also plausible that what we are seeing is actually two super-structures which are in-falling into each other along our line of sight. Such double super-structures may not be that unusual for the same reason, namely bias, that merging clusters are relatively common (e.g. \citealt{bvz}).

\

We will now discuss ways of reconciling the low probability of TONS08 intercepting the $z$=0.27 super-structure with the fact that we have detected it. It is possible, arguably probable, that two or more of these explanations are important.
\subsubsection{The value of $\sigma_{8h^{-1}}$}

These results are very dependent on the assumed normalisation of the power spectrum, $\sigma_{8h^{-1}}$. If we assume a lower value for $\sigma_{8h^{-1}}$ as preferred by combined 2dF/CMB measurements, then the expected number of super-structures in TONS08 decreases to 9$\times10^{-4}$. There is still much controversy over the value of $\sigma_{8h^{-1}}$. Recent work suggests that the higher values of $\sigma_{8h^{-1}}$ obtained from weak lensing techniques may be due to systematic errors and that the values are converging to lower values in line with studies of X-ray clusters and combined 2dF/CMB estimates. However, these results all assume $\Omega_{\rm{m}}$ = 0.3. Assuming a slightly lower value for $\Omega_{\rm{m}}$ would increase $\sigma_{8h^{-1}}$ to nearly 1. In all of this analysis, our results actually depend on $\sigma_{8h^{-1}} \times b$. We therefore favour $\sigma_{8h^{-1}}$=0.94 in this paper because it is closest to $\sigma_{8h^{-1}} \times b$ for optical galaxies and is a much better determined quantity than that of the dark matter.

\subsubsection{A different radio galaxy bias factor?}

Another explanation of our results is that radio galaxies are extremely biased tracers of mass in the $z$=0.27 super-structure (i.e. more biased than local radio galaxies). This could be due to redshift evolution of the bias factor or to different mechanisms being at work on larger scales, or a combination of the two.

An evolution of bias with redshift could be linked to the well known evolution in the space density of radio galaxies with redshift (e.g. \citealt{wil01}). \citet{cro} look at the evolution of QSO clustering as a function of redshift in the 2dF QSO redshift survey. They find that QSOs have similar clustering properties to local galaxies at all redshifts in their survey (z$\le$3.0). Because the clustering of dark matter is thought to decrease at higher redshifts (because fluctuations grow under gravity), this implies an increase in the bias factor of optically-selected QSOs at higher redshifts. The similar clustering properties of radio galaxies at low redshifts \citep{pn} and at $z\approx$1 \citep{bw} suggests similar results for radio galaxies. If the correlation length of radio galaxies is similar at high redshifts to that found locally, there must be an evolution in bias. Assuming a linear evolution of bias with redshift of the form $b(z) \propto (1+z)^n$, a value of $n$=0.4 is required for stable clustering in proper coordinates. Taking the local radio galaxies bias factor to be $b(0)$=1.65, this implies a radio galaxy bias factor $b(0.27)$=1.8 at $z$=0.27. This increase in the bias factor with redshift is fairly small at these relatively low redshifts and is also very model dependent. However, because the probability of TONS08 intercepting a super-structure is such a steep function of bias, only a small increase in bias may be enough to mean that the radio galaxies are tracing a less rare fluctuation in the evolved power spectrum that we would be more likely to intercept. We note here however that these calculations are based on data within a much larger survey area. A larger redshift evolution of bias may be possible if the triggering mechanism that causes a higher bias is peculiar to super-structures. 
We now discuss the possibility of a different radio galaxy bias for large scales. First, the $z$=0.27 super-structure looks to be an unusually dense region of sky. It is likely to be rich in collapsed substructure, e.g. rich clusters of galaxies (see also \citealt{hill}). The bias factor for rich clusters is $\approx$3.1 (\citealt{pd}; \citealt{pli}). This is precisely the value required to make the probability of seeing a super-structure in TONS08 $\approx$ 1 (Fig.~\ref{fig:prob_bias}) which would be the case if the radio galaxies are associated with rich clusters. Enhanced radio galaxy triggering due to group-group mergers (e.g. cluster formation) should also be considered \citep{sr}. Large surveys of radio galaxies would include few of these rare systems (i.e. clusters or forming clusters) and on average give lower values of bias (\citealt{pn}; \citealt{bw}) i.e. more appropriate to typical radio galaxy environments. 

The fact that the system may be collapsing may also be relevant. At recent epochs, structures the size of rich clusters are going non-linear. At least the central parts of super-structures have reached their turn-around radius \citep{gs} so perhaps this effect is peculiar to collapsing systems. Recent observations of large-scale structure in the 2dFGRS \citep{nor}, N-body simulations and cluster catalogues (\citealt{kbp}, \citealt{ein}) reveal that the matter in the Universe follows a filamentary structure. The largest structures lie at the nodes which join filaments and material falls into these along the filaments. Matching radio galaxies with 2dF galaxies reveals that radio galaxies are often on the edge of high density regions (Brand et al. in prep.) This is also seen for quasars \citep{scc}. In the TONS08 $z$=0.27 super-structure, only 2 of the 4 7CRS (the radio-brightest objects) are in the central region defined by a 50 Mpc radius sphere. Perhaps radio activity is triggered by the high velocities as the galaxies fall into the centre of the super-structure. Indeed, there are suggestions that nearby radio galaxies such as those in the Perseus cluster may have been triggered by their high velocities \citep{ped}. Note, also, that some degree of redshift space distortion is inevitable in a collapsing system which will inevitably and spuriously enhance the apparent overdensity of the super-structure (see Sec.~\ref{sec:zdistort}).

\subsubsection{Poisson sampling fluctuations}

Fig.~\ref{fig:prob_bias} also shows the 68 per cent confidence limits of the galaxy overdensities on the probability of TONS08 intercepting a super-structure as a function of bias. Poisson sampling fluctuations could mean that there happen to be more or fewer radio galaxies than we would expect on average for a fluctuation of a given size. This could modify our results slightly in either direction but not enough to explain our results. Fig.~\ref{fig:overdense} shows that this is a more severe problem for sparser samplers of the mass overdensity, such as radio galaxy surveys with higher radio flux density limits \citep{lac} or radio galaxy surveys at low redshifts at which the space density of radio galaxies is much lower \citep{pn}. 

\subsubsection{A representative volume?}

In the previous analysis, we have assumed that the TONS08 survey area is completely random. Although the original three 7CRS fields were essentially random in their position on the sky, our reason for choosing the TONS08 survey area was to follow up an overdensity already noticed in the 7C-II region. This obviously makes it more likely that we found an overdense region but by a factor that is hard to quantify until we have complete large-scale structure measurements in other TONS regions.

\subsection{The z=0.27 and z=0.35 super-structures}
\label{sec:discussion2}

If we again assume $\sigma_{8h^{-1}}$=0.94 and that the radio galaxy bias parameter $b$=1.65, Fig.~\ref{fig:prob_bias} forces us to conclude that the $z$=0.35 super-structure is an even rarer (5.1$\sigma$) fluctuation of which we would only expect $\sim$3.3E-4 in our sample volume. Assuming that it is real, the same arguments as made in Sec.~\ref{sec:discussion1} apply to the $z$=0.35 super-structure. Because it corresponds to a rarer fluctuation, and because there are two super-structures in the volume, many of the arguments are not strong enough. For example, a simple evolution of bias with redshift would have to be fairly extreme and would contradict other studies (e.g. \citealt{bw}).

An analysis of angular clustering in TONS08 (Sec.~\ref{sec:ang}) shows a possible edge in the distribution of radio galaxies within the $z$=0.35 super-structure. Its significance (Sec.~\ref{sec:sig}) is determined for the whole angular extent of TONS08. This will not take into account the fact the the $z$=0.35 super-structure is probably smaller than this and that number of radio galaxies expected in the same volume will also be smaller. Fig.~\ref{fig:overdense} shows that it is fairly unlikely that the overdensity is small enough to make the collection of radio galaxies not significant. For the following discussion, we therefore assume the super-structure to be real. 

Perhaps, the two fluctuations are linked. They are separated by about 310 Mpc and the power spectrum may still have power on these large scales \citep{mb}. The rarer the fluctuation, the more highly clustered it will be. Perhaps we observe these two rare fluctuations because they are sitting on top of an even rarer larger-scale fluctuation. The two-point correlation function is defined as the excess over Poisson of the joint probability of finding objects in two volume elements which are separated by a distance r. Using equation 3 from \citet{kai}, we determine that the correlation function for the super-structures is boosted by a factor of $\approx$300 over the typical value at this separation. Although boosted by a large amount, the correlation function of mass at these separations is very small ($\approx$0.001). The correlation function will only be boosted to a modest 0.5. Therefore, the probability of finding a second super-structure is 1.5 times higher than it would otherwise be (half of all super-structures separated by 300 Mpc are genuinely clustered and half are there by chance). This is a fairly negligible effect compared to the other uncertainties so we can effectively just multiply together the probabilities of finding each of the super-structures to obtain a joint probability of $\sim10^{-5}$. As we discussed in Sec.~\ref{sec:discussion1}, we may have targeted an especially dense region of the Universe because we have followed up a structure already discovered in the 7CRS \citep{rhw}. If this is the case, we may only be able to include the newly discovered $z$=0.35 super-structure in our analysis, bringing the joint probability down to $\sim 10^{-4}$.

We conclude that an increase in bias on the large scales of the super-structures and/or redshift space distortions are probably needed to explain finding two super-structures in the TONS08 region.

\subsection{Finding similar super-structures in other surveys}

\citet{gcc} combine the results of six quasar surveys to look for super-structures and find evidence for three structures. These are smaller and more elongated than the structures we have found. \citet{cow} find evidence for a huge structure $\sim 70 \times 140 \times 140$ Mpc$^3$ as traced by 18 quasars at redshift $z$=1.3. This is the largest structure at high redshifts yet found. 
Both the 2dFGRS \citep{col} and 2dFQSO surveys \citep{cro} should be able to find super-structures especially, if radio emission is a particularly biased tracer of mass, when combined with radio surveys such as NVSS or FIRST. The 2dFQSO survey has some evidence for clustering on large scales. However, they have not yet reported evidence for any super-structures. Radio loud quasars have been studied but only approx 2 per cent of optical and radio IDs match up meaning that the space density of radio loud quasars may not be sufficient at $z\approx$0.5 to trace super-structures. Similarly, the 2dFGRS team have not yet reported any newly discovered super-structures. Possible reasons include the fact that matching algorithms may cause incompletenesses, sky incompletenesses may prevent studies of the largest scales or it may be simply that noone has looked properly yet.

Sec.~\ref{sec:lacy} discusses the discovery of 2 redshift peaks in the \citet{lac} survey. The corresponding mass overdensities (assuming $b$=1.65) are so large that even though the survey volume is $\approx$ 7-times larger (i.e. 7-times more independent 50 Mpc radius spheres), the probability of there being two such super-structures in the survey are tiny ($\approx$ 1E-9 for the $z$=0.28 redshift peak). We should be cautious of this result for the following reasons. Because of the small number statistics, the errors are larger than for TONS08 and there is a good chance that the real overdensities are smaller (Fig.~\ref{fig:overdense}) and hence correspond to less rare primordial fluctuations. The log-normal approximation for the non-linear corrections we apply \citep{cj} are also known to break down for the highest peaks so the non-linear corrected overdensities may be inaccurate. Even if these super-structures are only as overdense as the TONS08 superstructures, the probability of intercepting two such structures is still fairly low. 
\citet{benn} finds a group of four radio galaxies at $z\approx$0.31 spanning $\approx$20 Mpc on the sky in a similar radio galaxy redshift survey. It seems that these huge structures in other surveys may not have looked very significant because the survey area is less well matched than TONS08 to super-structure scales and/or because of the effects of Poisson sampling fluctuations when the average space density of the radio galaxies are low. 

If an increase in bias is not due to a redshift evolution but instead due to some special triggering mechanism in collapsing super-structures then \citet{pn} should have detected similar super-structures. After all, even in a $\Lambda$CDM Universe, there are expected to be more super-structures of a given mass collapsing per unit volume at lower redshift. With a redshift limit of $z\approx$0.1 and sky area of 9.3 sr, their survey volume is naively of order  15-times larger than the volume probed by TONS08. They find 9 structures with 4 or more radio galaxies within a 57 Mpc radius sphere. The flux density limit of \citet{pn} is $S_{1.4}\approx$500 mJy with a redshift limit of $z\approx$0.1. If we compare this to TONS08, we find that the radio luminosity of the faintest radio galaxy detectable at the redshift limit at $z\approx$ 0.1 is $\approx 5 \times$ that of TONS08. By calculating the model redshift distribution for the survey limits of the \citet{pn} sample, we predict that there should be a total of 312 radio galaxies at $z<$0.1 (this compares well with the total number of 310 in the sample). With a co-moving volume of 2.25 $\times 10^8$ Mpc$^3$, we expect 290 independent spheres of radius 57 Mpc. We would therefore expect a mean number of $\sim$1.1 radio galaxy per 57 Mpc radius sphere. 
\citet{pn} only look for a minimum of 4 objects in each 50 Mpc radius sphere. As only 3.3 radio galaxies are needed to produce an overdensity of 2, the small number statistics involved at low redshifts ensure that many ($\approx$ 60 per cent of) overdensities of this size will not be detected. Using the method in Sec.~\ref{sec:bias}, and assuming the canonical bias of 1.65, we would expect \citet{pn} to have found $\approx$ 0.8 of these in their survey (assuming a redshift of $\approx z$=0.05).
Because they look for a minimum of 4 objects, \citet{pn} should only have been able to detect overdensities of 2.6 or more (neglecting Poisson uncertainties). Again assuming the canonical bias of 1.65, we would expect them to have found only 1.7E-2 of these in their survey volume. If we increase the bias to $b$=4, our model predicts that they should find 11 which is in good agreement with what they find.


The fact that similar structures appear in other surveys at both similar and lower redshift, reinforces the suggestion that we are seeing some mechanism at work which boosts the bias of the radio galaxies within super-structures. 

\subsection{Alternative structure formation theories}

The currently accepted theory of structure formation is that of the freezing in of quantum fluctuations of a scalar field during an inflationary period. An alternative theory that predicts larger numbers of rare fluctuations than the accepted inflationary theory is that large-scale structure is seeded by topological defects which are formed naturally during a symmetry breaking phase transition in the early Universe \citep{tur}. Although recent measurements of the position and amplitude of the acoustic peaks in the CMB \citep{beno} rule out cosmic defects as the dominant structure formation mechanism, it is possible that they provide a non-negligible contribution at large scales \citep{sak}. This would boost the number density of super-structures. 

\section{CONCLUDING REMARKS}

In this paper we have presented evidence of two significant super-structures in the TONS08 radio galaxy redshift survey as traced by radio galaxies in 50 Mpc radius spheres. From the implied overdensities of $\approx$ 3 on these large scales, we calculate the mass of these super-structures to be $\approx5\times10^{17} \rm{M_\odot}$. The super-structure at $z$=0.27 (traced by at least 13 radio galaxies) was tentatively identified in the 7CRS \citep{rhw}, traced by the brighter radio galaxies. The super-structure at $z$=0.35 (traced by at least 12 radio galaxies) is newly discovered.

Assuming $\sigma_{8h^{-1}}$=0.94 in a $\Lambda$CDM Universe and a radio galaxy bias factor of 1.65 these super-structures correspond to the evolved counterparts of $\sim 4$ and $5 \sigma$ fluctuations in the primordial density field. The expected number of such fluctuations in a volume equivalent to that of the TONS08 survey is $\approx$ 0.05 and $\approx 3 \times 10^{-4}$.

The probability of seeing two such fluctuations in TONS08 is strongly dependent on how representative the survey region is of the Universe. If, by targeting a structure discovered in the 7CRS, we have focused on a particularly unusual part of the Universe, then the probability of seeing two super-structures is roughly equal to the probability of finding only the $z$=0.35 super-structure, namely $\sim 10^{-4}$. If, however, the TONS08 target region turns out to be a reasonably representative one, then given our assumptions, the probability of seeing two super-structures is $\sim 10^{-5}$. The presence of similar radio galaxy overdensities (\citealt{lac}; \citealt{benn}) at similar redshifts suggests that these overdensities are actually fairly common. 

From Fig.~\ref{fig:prob_bias}, it is clear that the probability of detecting a radio-selected super-structure is a very strong function of the assumed radio galaxy bias. If we assume evolution of bias with redshift of the form $b(z) \propto (1+z)^n$, we would need $n \approx$ 2 to increase the probability of intercepting a super-structure such as the $z$=0.35 super-structure to a reasonable value. Previous studies suggest that this is an unfeasibly large evolution (\citealt{cro}; \citealt{bw}). Similar structures are also found by \citet{pn} at low redshifts, again suggesting that this cannot be a purely evolutionary effect.

Alternatively, an increase in bias could be due to some special triggering mechanism in collapsing super-structures. This seems the most promising way of reconciling the low probability of TONS08 intercepting both super-structures with the fact that we have detected them. There are several plausible explanations for this. Perhaps collapsed sub-structure within the super-structure, i.e. rich clusters, hosts the radio galaxies and the high bias simply reflects that of the rare rich clusters. However, similar super-structures have been found with radio quiet QSOs \citep{cow} and these tend to be found in less rich environments. 
Perhaps there is an enhanced rate of group-group mergers within the super-structure which is enhancing radio galaxy triggering. Perhaps as the super-structure goes non-linear, redshift space distortions and/or enhanced triggering due to high velocities boost the bias. 

We have just completed radio galaxy redshift surveys in other regions. These, along with radio galaxies selected from SDSS \citep{sto}, should determine how rare the TONS08 super-structures really are and, assuming structure formation models are correct, accurately determine the radio galaxy bias within super-structure regions. To determine the origin of this enhanced bias, we have begun a K-band (UKIRT) imaging survey of radio galaxies within the super-structures. By comparing structural parameters, merging rates and clustering environments with radio galaxies of similar optical magnitudes and radio flux densities out of super-structures, it should be possible to discriminate between the above possibilities. 

We conclude that despite the seemingly remarkable discovery of two rare, huge and massive super-structures in the fairly small volume of the TONS08 survey, this probably does not present any serious challenge to the standard $\Lambda$CDM model and the inflationary (i.e.\ Gaussian fluctuation) paradigm for structure formation. 

\section{ACKNOWLEDGEMENTS}

We thank John Peacock, Lance Miller, Hugo Martel, Catherine Heymans, Matthew Graham and Chris Blake for useful discussions and Devinder Sivia, Will Percival, Jenny Grimes and Steve Croft for use of code and software. We also thank Amanda Bauer and Filipe Abdalla for obtaining some of the spectra. We thank those involved in the original 7C search of this region: Katherine Blundell, Julia Riley, David Rossitter, Chris Willott, and Margrethe Wold.
We thank the staff at McDonald, NOT and the WHT for the  technical support they have provided. This research has made use of APM, NVSS, FIRST \& POSS-II web-based databases. 
The HET is operated by McDonald Observatory on behalf of The University of Texas at Austin, the Pennsylvania State University, Stanford University, Ludwig-Maximilians-Universitaet Muenchen, and Georg-August-Universitaet Goettingen.
The Nordic Optical Telescope is operated on the island of La Palma jointly by Denmark, Finland, Iceland, Norway, and Sweden, in the Spanish Observatorio del Roque de los Muchachos of the Instituto de Astrofisica de Canarias.
The WHT is operated on the island of La Palma by the Isaac Newton Group in the Spanish Observatorio del Roque de los Muchachos of the Instituto de Astrofisica de Canarias. This material is based in part upon work supported by the Texas Advanced Research Program under Grant No. 009658-0710-1999"
Kate Brand is supported by a PPARC studentship.

\clearpage

\begin{table*}
\vbox to400mm{\vfil 
\begin{picture}(0,0)
\put(460,-65){\includegraphics{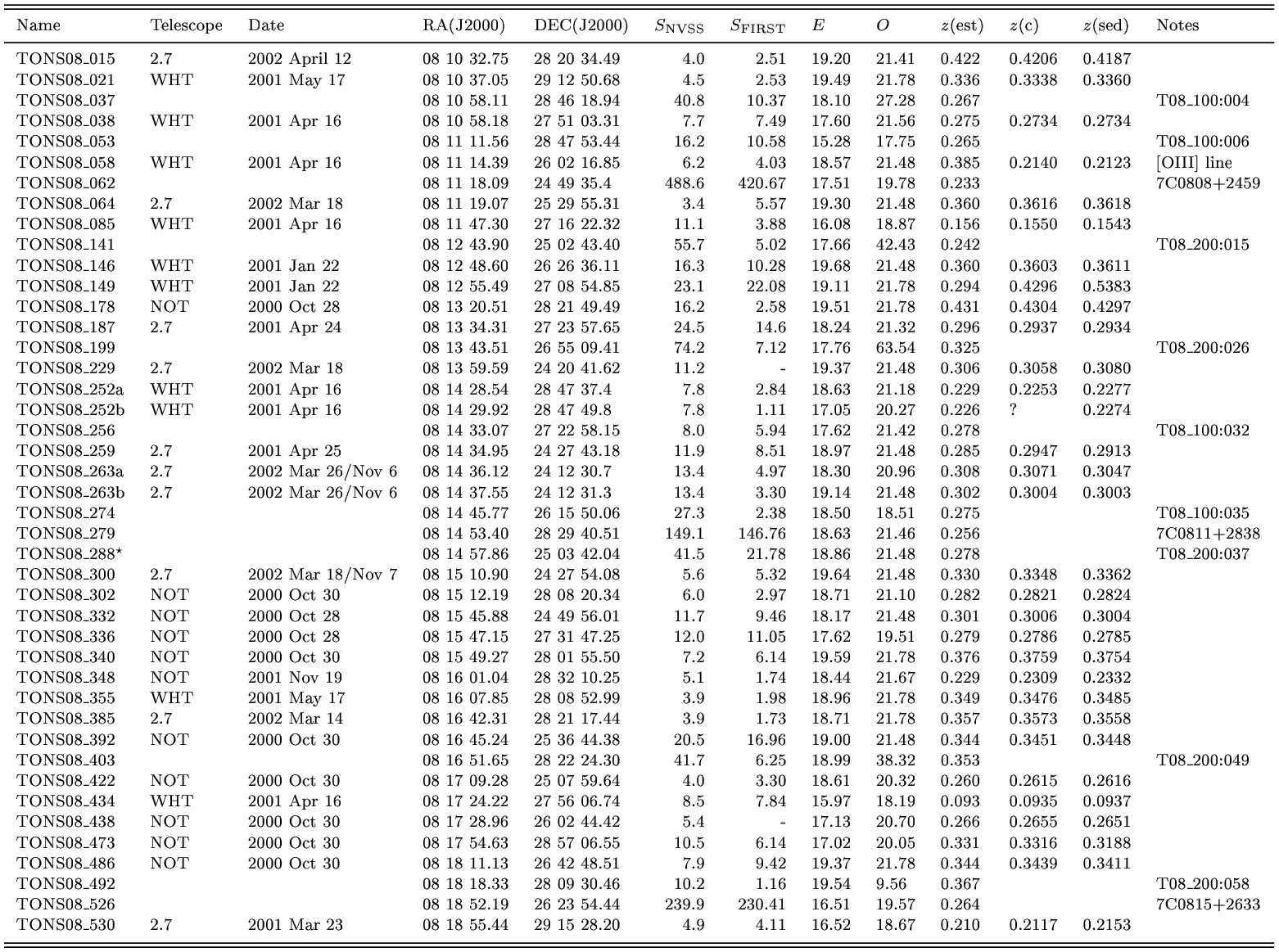}}
\end{picture}
{\caption[Table 1]{\label{tab:summary} A summary of key information 
on the TONS08 sample. Note that the RA and DEC are that of the radio positions except in the case of multiple sources where the positions are from APM.  $S_{\rm{NVSS}}$ and $S_{\rm{FIRST}}$ are the integrated radio flux densities (mJy) in the NVSS and FIRST catalogues respectively. In cases where FIRST resolves out multiple components, $S_{\rm{FIRST}}$ is the total integrated flux density. The $O$ and $E$ magnitudes are the APM corrected magnitudes. $z$(est) is the redshift estimated by identifying lines by eye and calculating the mean redshift. $z$(c) and $z$(sed) are the redshifts obtained by cross-correlating the spectra with the combined de-redshifted spectra and the GISSEL model respectively (Fig.~\ref{fig:template}). The notes denote the name of the object in other samples. A ? denotes that the cross-correlation was unsuccessful. Spectra of the TOOT objects will be presented in Rawlings et al. (in prep); Spectra for the 7CRS objects can be found in \citet{wil02}. The redshift of 3C200 is reported in \citet{lrl}. $^{\star}$ The NVSS entry for TONS08\textunderscore288 seems to be the confusion of two separate radio sources - the given redshift is for the ID of the brighter, double lobed radio source, but both sources have APM IDs. We are awaiting spectroscopic confirmation that the other object is at a similar redshift.

}}
\vfil}

\end{table*}

\clearpage

\addtocounter{table}{-1}

\begin{table*}
\vbox to400mm{\vfil 
\begin{picture}(0,0)
\put(460,-65){\includegraphics{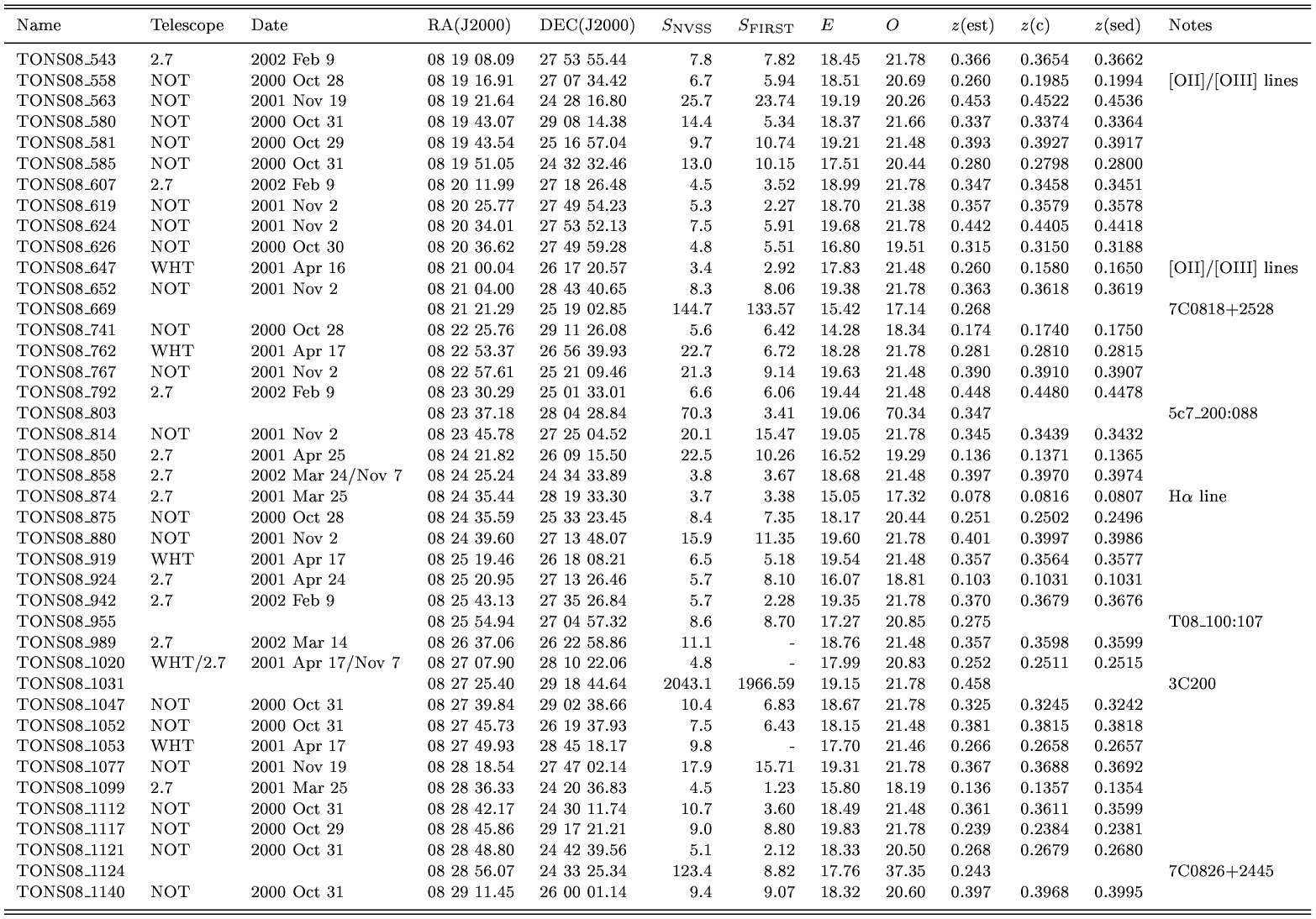}}
\end{picture}
{\caption{\bf (cont).}}
\vfil}
\end{table*}

\end{document}